\documentclass[]{aa}
\usepackage[varg]{txfonts}
\usepackage{graphicx}
\usepackage{amsmath}
\usepackage{amssymb}
\usepackage{lscape}
\usepackage{color}
\usepackage{longtable}
\usepackage{natbib,twoopt}
\usepackage{mathabx}
\usepackage{booktabs}
\usepackage{footmisc}

\begin{document} 

\title{Host galaxies of SNe Ic-BL with and without long gamma-ray bursts}

\author{
J. Japelj\inst{1}
\and
S. D. Vergani\inst{2} 
\and 
R. Salvaterra\inst{3}
\and 
M. Renzo\inst{1}
\and
E. Zapartas\inst{1}
\and
S. E. de Mink\inst{1}
\and
L. Kaper\inst{1}
\and
S. Zibetti\inst{4}
}

\institute{Anton Pannekoek Institute for Astronomy, University of Amsterdam, Science Park 904, 1098 XH Amsterdam, The Netherlands; \email{j.japelj@uva.nl}
\and GEPI - Observatoire de Paris Meudon. 5 Place Jules Jannsen, F-92195, Meudon, France
\and INAF - IASF Milano, via E. Bassini 15, I-20133, Milano, Italy
\and INAF-Osservatorio Astrofisico di Arcetri, Largo Enrico Fermi 5, I-50125 Firenze, Italy
}
 

\date{Received DD Mmmm YYYY / Accepted DD Mmmm YYYY} 

\abstract{Broad-line Ic supernovae (SNe Ic-BL) are a very rare class of core-collapse supernovae exhibiting high ejecta velocities and high kinetic energies. They are the only type of SNe that accompany long gamma-ray burst (GRB) explosions. Systematic differences found in the spectra of SNe Ic-BL with and without GRBs (GRB-SNe and SNe Ic-BL, respectively) suggest that either the progenitor or/and explosion mechanism of SNe Ic-BL with and without a GRB differ, or the difference could be only due to the viewing angle of the observer with respect to the orientation of the collimated explosion. We present the systematic comparison of the host galaxies of broad-lined SNe Ic with and without a detected GRB, the latter being detected in untargeted surveys, with the aim to find out whether there are any systematic differences between the environments in which these two classes of SNe preferentially explode. We study photometric properties of the host galaxies of a sample of 8 GRB-SNe and a sample of 28 SNe Ic-BL at $z < 0.2$. The two galaxy samples have indistinguishable luminosity and proper size distribution. We find indications that GRB-SNe on average occur closer to the centres of their host galaxies, i.e. the samples have a different distribution of projected offsets, normalized by the galaxy sizes. In addition we compare gas-phase metallicities of the GRB-SNe and SNe Ic-BL host samples and find that a larger fraction of super-solar metallicity hosts are found among the SNe Ic-BL without a GRB. Our results are indicative of a genuine difference between the two types of explosions and suggest that the viewing angle is not the main source of difference in the spectra of the two classes. We discuss the implications our results have on our understanding of progenitors of SNe Ic-BL with and without a GRB.}
  
 
\authorrunning{} 
\titlerunning{ }
\maketitle
  
\section{Introduction}

Massive stars can end their lives in very different ways. Following the collapse of their core and successful ejection of their outer layers \citep[e.g.,][]{Woosley2002}, the deaths of massive stars are marked by a variety of core-collapse supernova (SN) explosions \citep{Filippenko1997}. A subclass of core-collapse SNe are stripped-envelope supernovae, whose progenitor stars have lost most of the hydrogen (Type IIb, Ib and Ibn) or both the hydrogen and helium (Type Ic) in their outer envelopes prior to the collapse. Additionally, some supernovae of Type Ic are found to have very broad lines in their spectra (Type Ic-BL), indicative of very fast ejecta velocities \citep{Modjaz16}. The detection of Ic-BL SN1998bw following GRB\,980425 \citep{Galama1998} provided a link between the broad-lined supernovae and long gamma-ray burst explosions \citep[GRB;][]{Kumar2015}, which is now firmly established (\citealt{Hjorth2003,Malesani2004}, see e.g \citealt{Woosley2006,Cano2017} for a review). The subject of this paper are SNe Ic-BL with and without associated GRBs, in the following being referred to as GRB-SNe and SNe Ic-BL, respectively.

Mass, binarity, metallicity, rotation rate, mass-loss and probably magnetic fields play critical roles in forming evolved objects and determine the final fate of stars \citep[e.g., see reviews by][]{Maeder2000,Heger2003,Langer2012}. The fraction of broad-lined SNe Ic compared to the general population of core-collapse SNe is only $\sim 1\%$ \citep[e.g.,][]{Smith2011,Graur17}. Evidently the occurrence of a broad-line Ic SN demands special conditions that only a small fraction of massive stars satisfy. The fraction of long GRBs\footnote{In this paper we in general do not distinct between low and high luminosity GRBs linked to the GRB-SN \citep[e.g.][]{Soderberg2006}.} compared to the core-collapse SNe is likely even lower \citep{Soderberg2006,Guetta2007,Ghirlanda2013}. However, because the GRB explosions are highly collimated \citep{Granot2007,Ghirlanda2013,Eerten2018}, the GRB rate estimates come with large uncertainties as the typical beaming angles are poorly known. 

Due to the collimated GRB explosions it is also not straightforward to understand whether all SNe Ic-BL are linked to a GRB explosion or not: a non-detection of a GRB for a particular SN Ic-BL could simply be due to the explosion being directed away from us. Because these SNe occur at relatively large distances, it is hard to obtain direct evidence regarding their progenitors: until now only one SN Ib (iPTF13bvn; e.g. \citealt{Cao2013,Groh2013}) and potentially one SN Ic (SN2017ein; \citealt{vanDyk2017}) progenitor have been detected. 
Nevertheless, some evidence exists that at least some SNe Ic-BL are not accompanied by a GRB. For example, recent radio studies of large samples \citep{Corsi16} show that not all SNe Ic-BL are connected to GRBs (see also \citealt{Soderberg2006,Soderberg2010}). On the other hand polarimetry \citep[e.g.][]{Wang2008} and late-time spectroscopy \citep[e.g.][]{Taubenberger2009} of the SNe Ib/c indicate that all these explosions are to some degree aspherical. The difference between a SN with and without a GRB may therefore be a successful break-out of a jet from the star in the GRB case \citep{Lazzati2012,Barnes2017}. Supposing that there are two genuine classes of broad-line SNe, i.e. those with and without a GRB, it is still to be understood which are the factors leading to the separate outcomes. 

More information can be obtained by comparing the SN Ic-BL and GRB-SN population properties \citep[e.g.][]{Cano2017}. The number of detected and observed stripped-envelope SNe (including SNe Ic-BL) has increased in the recent years \citep[e.g.][]{Modjaz14,Modjaz16,Prentice16,Lyman2016} mostly thanks to the transient all-sky surveys such as the Palomar Transient Factory \citep{Law09}. The number of well studied GRB-SNe is also slowly but steadily rising \citep[e.g.][]{Kann2016,Cano2017}, allowing for a statistical comparison between the two populations. Comparing the optical spectra of SNe Ic-BL and GRB-SNe, \citealt{Modjaz16} showed that the two classes present clear differences, with the spectra of GRB-SNe having higher absorption velocities and broader line widths. This difference also shows itself in the higher kinetic energies of the GRB-SNe, as determined from the studies of their bolometric light curves \citep{Cano2013}.

Another line of evidence pointing to the nature of the progenitors lies in their host galaxy environment. For example, the so-called "collapsar" single-star evolution model, in which the progenitors of GRBs are massive stars with a high core rotation rate, indeed predicts that the progenitor star should have low metallicity in order to minimize the angular momentum losses and keep the high rotation rate of the core \citep[e.g.][]{Woosley2006}. However, the gas-phase metallicities found by analysing large samples of GRB host galaxies extend to higher values \citep[e.g.][]{Vergani2015,Kruhler2015,Japelj2016,Perley2016,Vergani2017} as predicted by this model \citep{Yoon2006}. Alternatively, GRBs could be formed in binary progenitor systems (e.g. formation through a common envelope \citep{Tout11,Zapartas17a}, tidal interaction \citep{Izzard04,Detmers08} or from runaway stars \citep{Cantiello07,Eldridge11}) which would imply the less stringent high-metallicity cut. Furthermore, clues about the progenitors are also found by investigating properties like star formation rate densities \citep{Kelly14} or ages of stellar populations in the hosts \citep[e.g.][]{Thoene2015}.

In particular, possible differences in the environments of GRB-SNe and SNe Ic-BL could reveal some properties of their progenitors. Any observed difference in this case cannot be attributed to the effect of the viewing angle. Several studies have focused on the host galaxies of stripped-envelope supernovae \citep[e.g.,][]{Modjaz08,Anderson10,Modjaz11,Sanders12,Kelly12,Kelly14,Graham2013}. In these studies the comparison either included the hosts of SNe detected in targeted surveys (which introduced a bias towards brighter galaxies) or the number of events of each class was low and the SNe Ic-BL with and without GRBs were combined into a single class. Broad-lined supernovae in general were found to prefer lower-metallicity environments than SNe Ib/c. Broad-line supernovae are furthermore found in galaxies with higher star-formation rate density with respect to SNe Ib/c \citep{Kelly14}. The interpretation of the results of the comparison studies is however difficult due to either the small sample size, to the inclusion of SNe from targeted surveys in the analysis or to the comparison of populations lying at completely different average redshift.

In this work we focus on the comparison of the morpho-photometric properties of a sample of untargeted SNe Ic-BL and GRB-SNe host galaxies at $z < 0.2$. The data and analysis are presented in Section \ref{selection}. In particular we compare the galaxy luminosities, sizes and explosion offsets from the host's centre (Section \ref{results}). Furthermore we collect available information of host galaxy spectra from the literature and measure gas-phase metallicities (Section \ref{metal}). In our study we find evidence that GRB-SNe are found at smaller offsets from galactic centres and in environments of lower metallicities. We discuss the implications of these results in Section \ref{progenitor}. Finally we provide conclusions in Section \ref{conclude}.

All errors are reported at 1$\sigma$ confidence unless stated otherwise. We use a standard cosmology \citep{Planck2014}: $\Omega_{\rm m} = 0.315$, $\Omega_{\Lambda} = 0.685$, and $H_{0} = 67.3$ km s$^{-1}$ Mpc$^{-1}$. All magnitudes are reported in the AB scale.

\section{Data and analysis}
\label{selection}

Supernovae can be discovered by targeted or untargeted surveys. Targeted surveys periodically look at a sample of pre-selected galaxies. This introduces a bias to the host galaxy population towards brighter and more metal-rich host galaxies on average \citep[e.g.][]{Modjaz11,Sanders12}. We therefore include only SNe discovered
by untargeted surveys in our samples. We searched through the available literature and The Open Supernova Catalog archive\footnote{https://sne.space/} \citep{Guillochon17} for all spectroscopically confirmed SNe Ic-BL detected in untargeted surveys. We require that the classification of the SNe is robust. For example, we do not consider PTF12gzk, a SN that shows high expansion velocities (similar to SNe Ic-BL), yet its spectra do not show the persistent broad lines that are typical for Ic-BL SNe \citep{BenAmi2012} and the automatic classification of \citet{Modjaz14} classifies it as a SN Ic. We limit our comparison to events occurring at $z < 0.2$, in order to have a similar redshift distribution of the SNe Ic-BL with and without a GRB. Since the southern part of the sky has not been covered by deep optical surveys at the time of our study, we limit our sample to declinations $\delta > -30\degr$. 29 SNe Ic-BL satisfy our selection criteria. However, the host of SN2014ad lies near a very bright star whose contamination in the images of the host is very hard to properly account for. Therefore we do not consider this one in the further analysis and we are left with a sample of 28 SNe Ic-BL. Eight spectroscopically confirmed GRB-SNe have been detected up to this redshift. The two samples are summarized in Table \ref{tab1}.

\begin{figure*}[!t]
\centering
\includegraphics[scale=0.58]{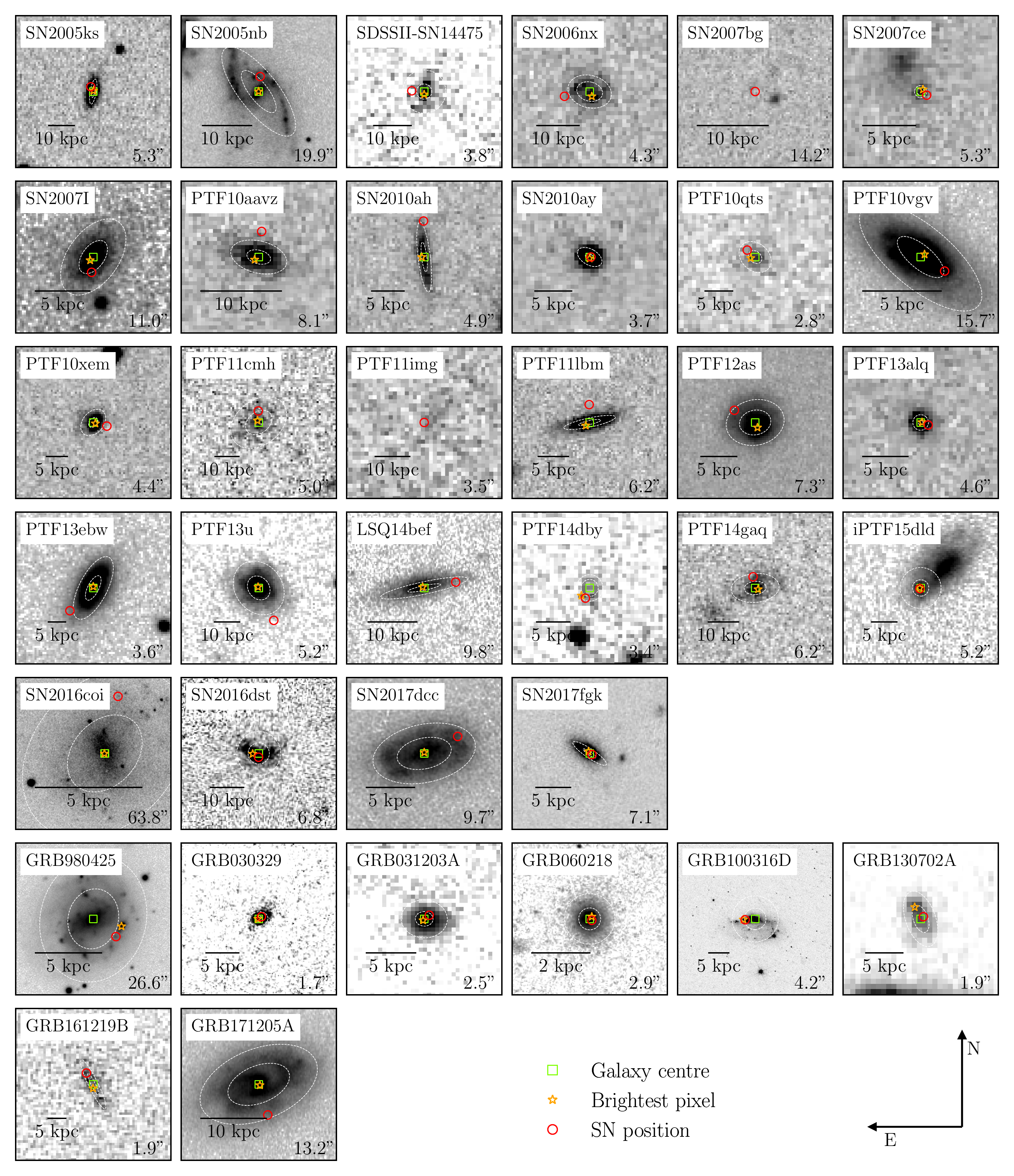}
\caption{Overview of the galaxies in the sample. See Table 1 for the information on the survey or the instrument with which each image has been taken. The red circle indicates the position of each SN: the size of the circle is arbitrary. The green square and the orange star indicate the measured galaxy centre and the brightest pixel in the galaxy, respectively. Overplotted are ellipses corresponding to the measured $r_{50}$ and $r_{90}$ radii, where the values of position angle and inclination (see Table \ref{tab6}) have been used to take geometric effects into account. The values provided in the bottom right corners correspond to the physical scales in each figure. All images are centred on galaxy centres except for the case of SN2016coi. Note that in the case of PTF11img the host galaxy is not detected.}
\label{fig1}
\end{figure*}

Our main resource of photometric images are deep all-sky surveys. In particular we use the Sloan Digital Sky Survey (SDSS) Data Release 12 \citep{Alam15} and PanSTARRS Data Release 1 \citep{Chambers16} data. For GRB-SNe deep exposures obtained via GRB follow-up studies are available. We collected the available images obtained with various facilities. We limit our imaging analysis to the data taken with the Sloan {\it r} filter (in the case of SDSS and PanSTARRS data) or to the images taken with filters with a similar wavelength range. The images of the galaxies in the sample are given in Figure \ref{fig1}. The detailed account is provided in Table \ref{tab1}. We search for the most accurate positions of detected SNe in the papers and, if those are not available, in the discovery telegrams. We note that for SNe Ic-BL the positions are quoted without uncertainty in the majority of cases. We discuss how this impacts our results in Section \ref{offsets}. The SN position coincides with a galaxy in nearly all the cases in our sample. In two cases we could not carry out all the measurements either because a galaxy is not detected (PTF11img) or because the galaxy is detected but too faint (SN2007bg) - we discuss the implications for the results later on. Several cases which merit special attention are discussed in Section \ref{individual}.

\begin{table*}
\small
\renewcommand{\arraystretch}{1.2}
\begin{center}
\begin{tabular}{lccccr}
\hline
\hline
SN/GRB      & Redshift  & Image        & RA$_{\rm SN}$   & Dec$_{\rm SN}$ & Ref \\
            &           &              & (hh:mm:ss)      & (dd:mm:ss)     &      \\
\hline
SN2005ks    & 0.0988    & SDSS-r       & 21:37:56.56     & -00:01:56.90   & \citet{Zheng2008}\\
SN2005nb    & 0.024     & SDSS-r       & 12:13:37.61     & +16:07:16.20   & \citet{Quimby2006} \\
SDSS-IISN14475 & 0.1440 & SDSS-r       & 22:24:30.96     & +00:12:12.28   & \citet{Sako2014}\\
SN2006nx    & 0.125     & SDSS-r       & 03:33:30.63     & -00:40:38.24   & \citet{Bassett2006} \\
SN2007bg    & 0.034     & SDSS-r       & 11:49:26.18     & +51:49:21.8    & \citet{Moretti07} \\
SN2007ce    & 0.046     & SDSS-r       & 12:10:17.96     & +48:43:31.51   & \citet{Quimby2007} \\
SN2007I     & 0.0216    & SDSS-r       & 11:59:13.15     & -01:36:18.9    & \citet{Jin07}\\
PTF10aavz   & 0.062     & SDSS-r       & 11:20:13.36     & +03:44:45.20   & \citet{Arcavi2010} \\
SN2010ah    & 0.0498    & SDSS-r       & 11:44:02.99     & +55:41:27.60   & \citet{Ofek2010} \\
SN2010ay    & 0.0671    & SDSS-r       & 12:35:27.19     & +27:04:02.78   & \citet{Drake2010}\\
PTF10qts    & 0.0907    & SDSS-r       & 16:41:37.60     & +28:58:21.10   & \citet{GalYam2010}\\
PTF10vgv    &  0.015    & SDSS-r       & 22:16:01.17     & +40:52:03.30   & \citet{Corsi12}  \\
PTF10xem    & 0.0567    & SDSS-r       & 01:47:06.88     & +13:56:28.80   & \citet{Corsi16} \\
PTF11cmh    & 0.1055    & SDSS-r       & 13:10:21.74     & +37:52:59.60   & \citet{Corsi16} \\
PTF11img    & 0.158     & SDSS-r       & 17:34:36.30     & +60:48:50.60   & \citet{Corsi16}\\
PTF11lbm    & 0.039     & SDSS-r       & 23:48:03.20     & +26:44:33.50   & \citet{Corsi16} \\
PTF12as     & 0.033     & SDSS-r       & 10:01:34.05     & +00:26:58.40   & \citet{Corsi16} \\
PTF13alq	& 0.054     & SDSS-r       & 11:48:02.03     & +54:34:38.60   & The Open Supernova Catalog \\
PTF13ebw    & 0.069     & SDSS-r       & 08:17:15.88     & +56:34:41.60   & \citet{Corsi16}\\
PTF13u      & 0.1       & SDSS-r       & 15:58:51.21     & +18:13:53.10   & \citet{Corsi16} \\
LSQ14bef    & 0.05      & PanSTARRS-r  & 14:27:41.77     & -08:36:45.50   & \citet{Polshaw14} \\
PTF14dby    & 0.074     & SDSS-r       & 15:17:06.29     & +25:21:11.40   & \citet{Prentice16}\\
PTF14gaq    & 0.0826    & SDSS-r       & 21:32:54.08     & +17:44:35.60   & \citet{Corsi16}\\
iPTF15dld   & 0.047     & SDSS-r       & 00:58:13.28     & -03:39:50.29   & \citet{Singer2015} \\
SN2016coi   &  0.003646 & SDSS-r       & 21:59:04.12     & +18:11:10.72   & Gaia Alerts$^{(a)}$ \\  
SN2016dst   & 0.074     & PanSTARRS-r  & 04:55:54.68     & -29:34:13.73   & Gaia Alerts \\
SN2017dcc   & 0.0245    & PanSTARRS-r  & 12:49:04.89     & -12:12:22.57   & Gaia Alerts \\
SN2017fgk   & 0.034     & PanSTARRS-r  & 17:47:49.19     & +16:08:05.18   & Transient Name Server$^{(b)}$\\
\hline
GRB\,980425 & 0.0088    & VLT/FORS2-R  & 19:35:03.31     & -52:50:44.7    & \citet{Tinney1998}\\
GRB\,030329  & 0.1685   & HST/ACS-F606W  & 10:44:49.96   & +21:31:17.44   & \citet{Taylor2004}\\
GRB\,031203A & 0.1055   & VLT/FORS2-R  & 08:02:30.16     & -39:51:03.51   & \citet{Soderberg04}\\
GRB\,060218 & 0.03302   & HST/ACS-F625W & 03:21:39.68     & +16:52:01.82   & \citet{Soderberg2007}\\
GRB\,100316D & 0.0593   & HST/WCS-F625W & 07:10:30.56     & -56:15:20.18   & \citet{Starling11} \\
GRB\,130702A & 0.145    & VLT/FORS2-R  & 14:29:14.77     & +15:46:26.37   & \citet{Singer2013}\\
GRB\,161219B & 0.1475   & PanSTARRS-r  & 06:06:51.42     & -26:47:29.52   & \citet{Alexander16}\\
GRB\,171205A & 0.0368   & PanSTARRS-r  & 11:09:39.52     & -12:35:18.48   & \citet{Laskar17}\\
\hline
\hline
\end{tabular}
\end{center}
\caption{Samples of SNe Ic-BL without (28) and with (8) a GRB studied in this work. For each host galaxy we report: redshift, survey or instrument (and filter) of the images that were analysed, and the position at which a SN occurred. References for positions are given in the last column. Images of hosts of GRB-SNe, which are not part of all-sky surveys, were retrieved from: GRB\,980425 (ESO archive, \citealt{Sollerman2005}), GRB\,030329 (HST archive, \citealt{Sollerman2005}), GRB\,031203A (ESO archive, \citealt{Mazzali2006}), GRB\,060218 (HST archive, PI: S.R. Kulkarni), GRB\,100316D (HST archive, \citealt{Cano2011}), GRB\,130702A (ESO archive, PI: E. Pian). \newline (a) $http://gsaweb.ast.cam.ac.uk/alerts$ \newline (b) $https://wis-tns.weizmann.ac.il/$}
\label{tab1}
\end{table*}

\begin{table*}
\tiny
\renewcommand{\arraystretch}{1.2}
\begin{center}
\begin{tabular}{lcccccccr}
\hline
\hline
SN/GRB      & RA$_{\rm host}$  & Dec$_{\rm host}$      & $m_{\rm R}$   & $M_{\rm R/(1+z)}$ & $r_{50}$      & $r_{90}$     & Offset$_{\rm centre}$& Offset$_{bp}$\\
            &                  &                       & (mag)         & (mag)             & (kpc)         & (kpc)        & (kpc)                & (kpc) \\
\hline
SN2005ks    &  21:37:56.54     & -00:01:57.84          & 18.48 +- 0.17 & -19.92 +- 0.17    & 2.26 +- 0.09  & 4.76 +- 0.19 & 1.88                 & 1.74\\
SN2005nb$^{(a)}$& 12:13:37.65  & +16:07:10.29          & 14.66 +- 0.01 & -20.58 +- 0.01    & 5.13          & 10.84        & 2.99                 & 2.97\\
SDSS-IISN14475 & 22:24:30.87   & +00:12:12.22          & 20.93 +- 0.12 & -18.36 +- 0.12    & 1.97 +- 0.21  & 5.59 +- 0.71 & 3.54                 & 3.33\\
SN2006nx    &  03:33:30.47     & -00:40:37.81          & 19.96 +- 0.06 & -19.11 +- 0.06    & 3.00 +- 0.14  & 5.07 +- 0.37 & 5.80                 & 6.22\\
SN2007bg    &  11:49:26.24     & +51:49:22.71          & > 22.4        & > -13.6           &               &              &                      & \\
SN2007ce$^{(b)}$& 12:10:18.02  & +48:43:31.87          & 20.85 +- 0.24 & -15.78 +- 0.24    & 0.34 +- 0.02  & 0.65 +- 0.23 & 0.65                 & 0.63$^{(c)}$\\
            & 12:10:18.19      & +48:43:34.69          & 19.33 +- 0.14 & -17.30 +- 0.14    & 2.48 +- 0.24  & 5.10 +- 0.67 & 3.67                 & \\ 
SN2007I     &  11:59:13.13     & -01:36:15.88          & 17.40 +- 0.02 & -17.59 +- 0.02    & 1.79 +- 0.03  & 4.08 +- 0.08 & 1.38                 & 1.09\\
PTF10aavz   &  11:20:13.38     & +03:44:42.63          & 19.24 +- 0.04 & -18.13 +- 0.04    & 1.49 +- 0.07  & 3.51 +- 0.21 & 3.17                 & 3.55\\
SN2010ah    &  11:44:02.98     & +55:41:22.34          & 19.07 +- 0.18 & -17.71 +- 0.18    & 2.21 +- 0.06  & 5.21 +- 0.16 & 5.32                 & 5.34\\
SN2010ay    &  12:35:27.20     & +27:04:02.76          & 19.05 +- 0.03 & -18.40 +- 0.30    & 0.33 +- 0.01  & 1.07 +- 0.16 & 0.17                 & 0.25\\
PTF10qts    &  16:41:37.53     & +28:58:20.36          & 21.23 +- 0.21 & -16.91 +- 0.21    & 1.28 +- 0.26  & 3.07 +- 0.81 & 2.07                 & 1.51\\
PTF10vgv    &  22:16:01.59     & +40:52:06.07          & 15.29 +- 0.04 & -19.21 +- 0.04    & 1.81 +- 0.05  & 4.75 +- 0.19 & 1.75                 & 1.64\\
PTF10xem    &  01:47:07.06     & +13:56:29.48          & 18.60 +- 0.03 & -18.57 +- 0.03    & 1.26 +- 0.06  & 3.75 +- 0.17 & 3.18                 & 2.66\\
PTF11cmh    &  13:10:21.73     & +37:52:57.33          & 19.89 +- 0.33 & -18.55 +- 0.33    & 5.79 +- 0.56  & 12.25 +- 1.59& 4.54                 & 3.84\\
PTF11img    &                  &                       & > 22.5        & > -16.9           &               &              &                      & \\
PTF11lbm    &  23:48:03.18     & +26:44:30.05          & 17.95 +- 0.03 & -18.42 +- 0.03    & 1.79 +- 0.06  & 4.18 +- 0.22 & 3.62                 & 3.36\\
PTF12as     &  10:01:33.78     & +00:26:56.04          & 16.47 +- 0.01 & -19.44 +- 0.01    & 2.14 +- 0.01  & 3.95 +- 0.03 & 3.25                 & 3.90\\
PTF13alq	&  11:48:02.12     & +54:34:38.87          & 19.97 +- 0.10 & -17.00 +- 0.10    & 0.26 +- 0.04  & 0.84 +- 0.04 & 0.86                 & 0.74\\
PTF13ebw    &  08:17:15.32     & +56:34:46.12          & 16.27 +- 0.06 & -21.35 +- 0.06    & 3.66 +- 0.19  & 9.33 +- 0.74 & 8.83                 & 9.08\\
PTF13u      &  15:58:51.42     & +18:13:59.60          & 17.02 +- 0.02 & -21.38 +- 0.02    & 4.83 +- 0.10  & 10.84 +- 0.21& 13.56                & 14.0\\
LSQ14bef    &  14:27:42.24     & -08:36:46.45          & 17.52 +- 0.02 & -19.35 +- 0.01    & 3.34 +- 0.03  & 8.94 +- 0.80 & 7.14                 & 6.69\\
PTF14dby    &  15:17:06.28     & +25:21:11.90          & 22.14 +- 0.08 & -15.61 +- 0.08    & 1.23 +- 0.23  & 1.91 +- 0.35 & 0.76                 & 0.69\\
PTF14gaq    &  21:32:54.05     & +17:44:33.50          & 19.33 +- 0.20 & -18.80 +- 0.20    & 3.50 +- 0.25  & 7.66 +- 0.58 & 3.62                 & 4.30\\
iPTF15dld$^{(b)}$& 00:58:13.27 & -03:39:50.14          & 18.27 +- 0.12 & -18.43 +- 0.12    & 1.76 +- 0.07  & 4.07 +- 0.24 & 0.20                 & 0.12$^{(c)}$\\
            &  00:58:12.95     & -03:39:46.15          & 16.67 +- 0.07 & -20.03 +- 0.07    & 3.49 +- 0.14  & 7.57 +- 0.32 & 6.17                 & \\
SN2016dst   &  04:55:54.68     & -29:34:12.99          & 18.55 +- 0.04 & -19.11 +- 0.04    & 3.07 +- 0.07  & 6.33 +- 0.17 & 1.08                 & 1.94\\            
SN2016coi   &  21:59:04.69     & +18:10:36.67          & 12.95 +- 0.06 & -18.30 +- 0.06    & 2.00 +- 0.16  & 4.41 +- 0.40 & 2.70                 & 2.72\\  
SN2017dcc   &  12:49:05.34     & -12:12:25.91          & 15.90 +- 0.01 & -19.40 +- 0.01    & 2.79 +- 0.02  & 5.27 +- 0.07 & 3.72                 & 3.77\\
SN2017fgk   &  17:47:49.25     & +16:08:05.79          & 17.33 +- 0.03 & -18.81 +- 0.03    & 1.08 +- 0.03  & 2.80 +- 0.07 & 0.75                 & 0.62\\
\hline
GRB\,980425  & 19:35:04.00     &  -52:50:37.68         & 14.37 +- 0.05 & -18.66 +- 0.05    & 2.25 +- 0.03  & 4.74 +- 0.03 & 2.25                 & 0.87\\
GRB\,030329  & 10:44:49.94     &  +21:31:17.35         & 23.09 +- 0.02 & -16.42 +- 0.02    & 0.54 +- 0.03  & 1.37 +- 0.08 & 0.86                 & 0.59\\
GRB\,031203A & 08:02:30.18     &  -39:51:03.68         & 20.65 +- 0.03 & -20.19 +- 0.03    & 1.04 +- 0.04  & 2.79 +- 0.16 & 0.34                 & 0.64\\
GRB\,060218  & 03:21:39.69     &  16:52:01.88          & 20.30 +- 0.06 & -15.88 +- 0.06    & 0.37 +- 0.01  & 0.85 +- 0.05 & 0.08                 & 0.12\\
GRB\,100316D & 07:10:30.31     &  -56:15:20.21         & 17.80 +- 0.05 & -19.61 +- 0.05    & 2.55 +- 0.02  & 5.42 +- 0.10 & 2.48                 & 0.39  \\
GRB\,130702A & 14:29:14.78     &  +15:46:26.25         & 23.34 +- 0.12 & -15.88 +- 0.12    & 1.85 +- 0.16  & 3.77 +- 0.30 & 0.49                 & 1.64 \\
GRB\,161219B & 06:06:51.37     &  -26:47:30.62         & 20.67 +- 0.03 & -18.56 +- 0.03    & 3.19 +- 0.16  & 7.17 +- 0.62 & 3.45                 & 4.37\\
GRB\,171205A & 11:09:39.69     &  -12:35:11.31         & 15.33 +- 0.02 & -20.90 +- 0.02    & 4.86 +- 0.03  & 9.25 +- 0.06 & 5.77                 & 5.58\\
\hline
\hline
\end{tabular}
\end{center}
\caption{Results of the analysis of the host galaxies: barycentre position (RA$_{\rm host}$,Dec$_{\rm host}$) of the host galaxy, apparent and absolute magnitude, radii $r_{50}$ and $r_{90}$ (radii containing 50 and 90$\%$ of the light from the galaxy) and the projected offset of an explosion from the centre of the galaxy (offset$_{\rm centre}$) and from the brightest pixel (offset$_{\rm bp}$). Absolute magnitudes are corrected for Galactic foreground extinction \citep{Schlafly2011}, while the apparent magnitudes are not. Systematic errors on measured offsets are discussed in Section \ref{offsets}. \newline Notes: \newline(a) The surface brightness of this host can not be modelled with a simple phenomenological model, therefore the results are from photometric analysis and the measured characteristic radii are without errors.\newline(b) The first line corresponds to the measurement assuming the host galaxy is only the bright (blue) region near the SN explosion. The second line is obtained in the case the host galaxy is the whole complex. \newline(c) The brightest pixel is located in the compact region of the complex.}
\label{tab2}
\end{table*}

\subsection{Measuring luminosities and sizes}

The main parameters that we are interested in are galaxy luminosity, their (physical) size and the position of the SN site relative to the host galaxy centre (projected offset, in the following only offset). Galaxies in the sample span a large interval of apparent sizes, magnitudes and morphologies. The images that we analyse come from different instruments. Taking that into account, one has to do some compromises with the analysis (i.e. sometimes one technique gives more reliable result than the other). All galaxies in the sample have been analysed using SExtractor (v2.19.5; \citealt{Bertin1996}). For the measured apparent magnitudes we adopt the MAG$\_$AUTO magnitudes, corresponding to the flux within 2.5 \citet{Kron1980} radius. With SExtractor we can also obtain the measurements of $r_{50}$ and $r_{90}$, representing the radii within which 50 and 90$\%$ of the light is enclosed, respectively. However, many of the galaxies in the sample have a very small apparent size, comparable to the size of the point-spread function (PSFs) of the image. To take the PSF accurately into account, we measure galaxy sizes by fitting their surface brightness profile with a simple parametric model using GALFIT \citep{Peng2010}. We generate PSFs of individual images using the PSFEx \citep{Bertin2011}, a routine that together with SExtractor extracts models of the PSF at desired positions in the images. This PSF is then used as an input of the GALFIT fitting procedure. For simplicity we fit each galaxy with one Sersic profile and, if statistically justified, with an additional exponential profile with a common centre. Most of the galaxies in the sample are well described with this simple model. The exception are galaxies which are well resolved and very big on the sky, where the simple model cannot account for the presence of features like spiral arms. In these few cases (GRB980425, SN2005nb, SN2016coi) the PSF does not play an important role and we adopt the SExtractor measurements. We note that the magnitudes measured from the GALFIT analysis are consistent with those obtained with SExtractor. Measured values of $r_{50}$ are consistently smaller with the GALFIT for objects with $r_{50} \lesssim {\rm FWHM(PSF)}$, which is exactly what we expect.

With SExtractor we estimate the barycentre of each galaxy (simply centre in the following) and the brightest pixel in each galaxy (see e.g. \citealt{Lyman2017}). Finally we measure the offset between the SN position and the centre of their host galaxy. The measured positions - galaxy centre, the position of the brightest pixel and the position of the detected SN explosion - are indicated in Figure \ref{fig1} for each individual case.

The measured $r_{50}$ and $r_{90}$ radii are transformed to proper sizes and the magnitude is transformed to absolute magnitude as $M_{R/(1+z)} = m_{\rm R} - A_{\rm R} - 5\log\left( D_{\rm L}/10 {\rm pc} \right) + 2.5\log\left(1 + z\right)$, where $A_{\rm R}$ is the Galactic foreground extinction taken from the maps of \citet{Schlafly2011}. All the measurements are collected in Table \ref{tab2}.

\begin{figure}[]
\centering
\includegraphics[scale=0.58]{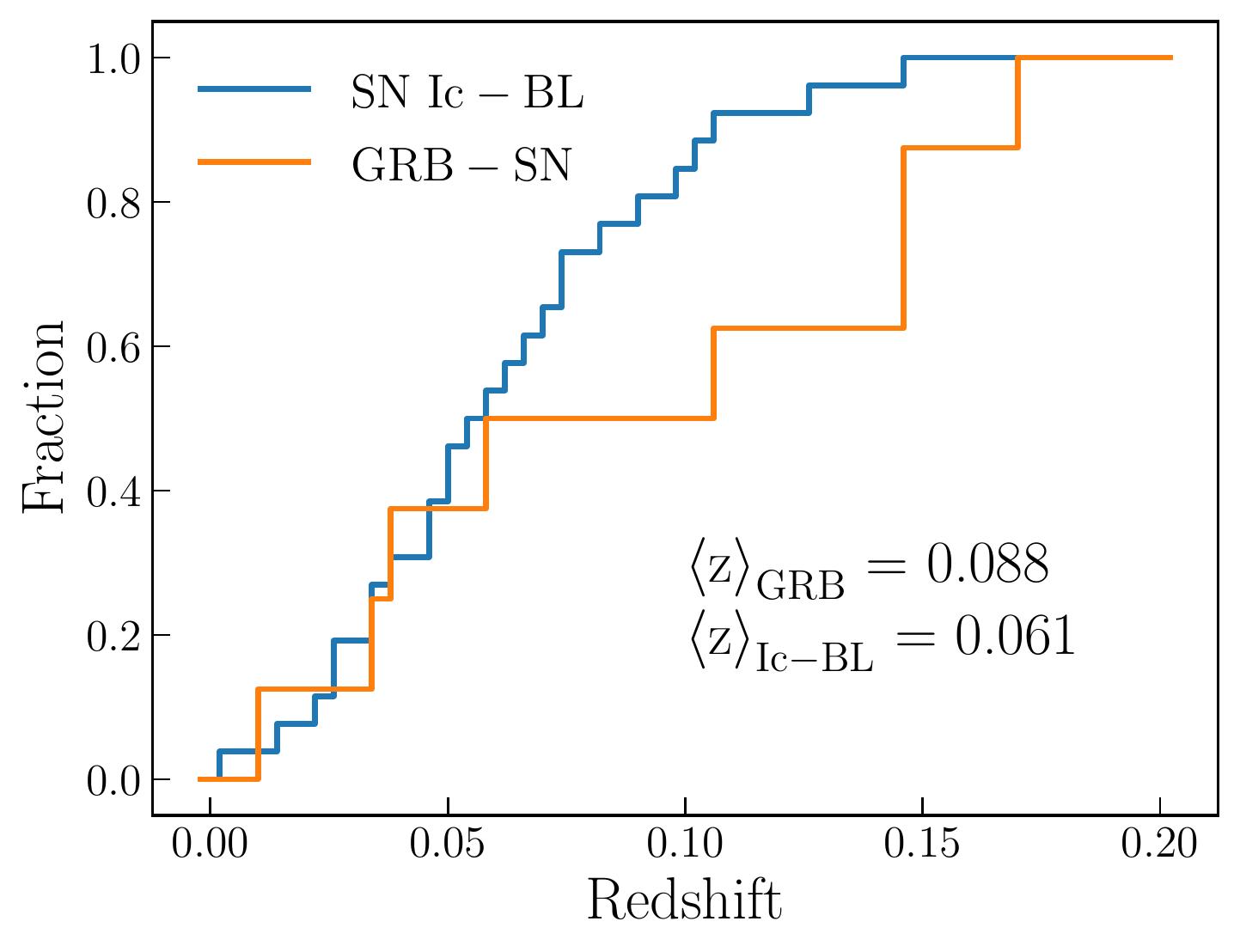}
\caption{Cumulative redshift distributions of the GRB-SN and SN Ic-BL samples. Indicated are mean redshifts of the distributions.}
\label{fig2}
\end{figure}

\subsection{Individual host galaxies}
\label{individual}

{\it SN2007bg:} There is a very faint source near the position of the SN, which we assume to be the actual host galaxy. The other two sources nearby are offset by such an amount that the projected offsets, normalized by the size of the galaxy (i.e. $r_{50}$), when computed, are in a completely different regime as for the rest of the sample. The presumable host is too faint to allow for a reliable magnitude and size measurement. We only provide a magnitude upper limit and an estimate of the projected offset. We note that \citet{Kelly12} presumably assumed the nearby bright source to be the host galaxy. However, if we make the same assumption, we measure a larger normalized projected offset as the one provided by \citet{Kelly12}.

{\it SN2007ce:} The SN occurred near the position of a compact, blue region. It is unclear whether the region is a part of the larger complex (see Figure \ref{fig1}). We conduct the analysis for both scenarios (see Table \ref{tab2}). For subsequent analysis we conservatively assume that the compact region is a galaxy on its own. 

{\it iPTF15dld:} The SN occurred at the position of a bright, compact starburst region (see Fig. \ref{fig1}). It is unclear whether the region is a part of the whole system or not (see also \citealt{Pian2017}). We conduct the analysis for both scenarios (see Table \ref{tab2}). For subsequent analysis we conservatively assume that the compact region is a galaxy on its own. In the alternative scenario the SN is found at a much larger normalized offset. 

{\it SN2005ks:} The reported position of the SN differs in \citet{Zheng2008} and \citet{Sako2014}, even though the position in both works is determined from the SDSS detection. We adopt the position determined by \citet{Zheng2008}. This position places the SN to about half the distance to the host's centre than the other one. 

{\it PTF11img:} The host galaxy is not detected in the SDSS image. We only derive a magnitude upper limit.

{\it GRB\,161219B:} An image of the host has been taken with the HST/WFC3 camera using the {\it F200LP} filter \citep{Cano2017b}. Because this filter covers much broader wavelength range compared to the other filters in this study, we prefer to use the PanSTARRS image for the analysis. We note that the host is much better resolved in the HST image. The analysis of the HST image results in very similar offsets and slightly larger sizes ($r_{\rm 50} = 3.75 \pm 0.08$ kpc) compared to the values measured from the PanSTARRS image.

\section{Results}
\label{results}

We compare the distributions of the studied characteristics of the two samples: galaxy size $r_{\rm 50}$ and $r_{\rm 90}$ and the offsets of SN location with respect to the galaxy centre and the brightest pixel. For each distribution we compute the median and 1$\sigma$-equivalent confidence interval - we do not take the errors in the computation of these statistics into account. We use the two-sided Anderson-Darling (AD) test between any two distributions to estimate a chance probability that the two distributions of a given characteristic are drawn from the same parent distribution. The errors are taken into account by performing a Monte Carlo (MC) simulation: for 10000 times we perform the AD test on the distributions built by randomly varying the measured characteristic for its error. The results are summarized in Table \ref{tab5}. 

\subsection{Absolute magnitude and size}

\begin{figure}[]
\centering
\begin{tabular}{c}
\includegraphics[scale=0.55]{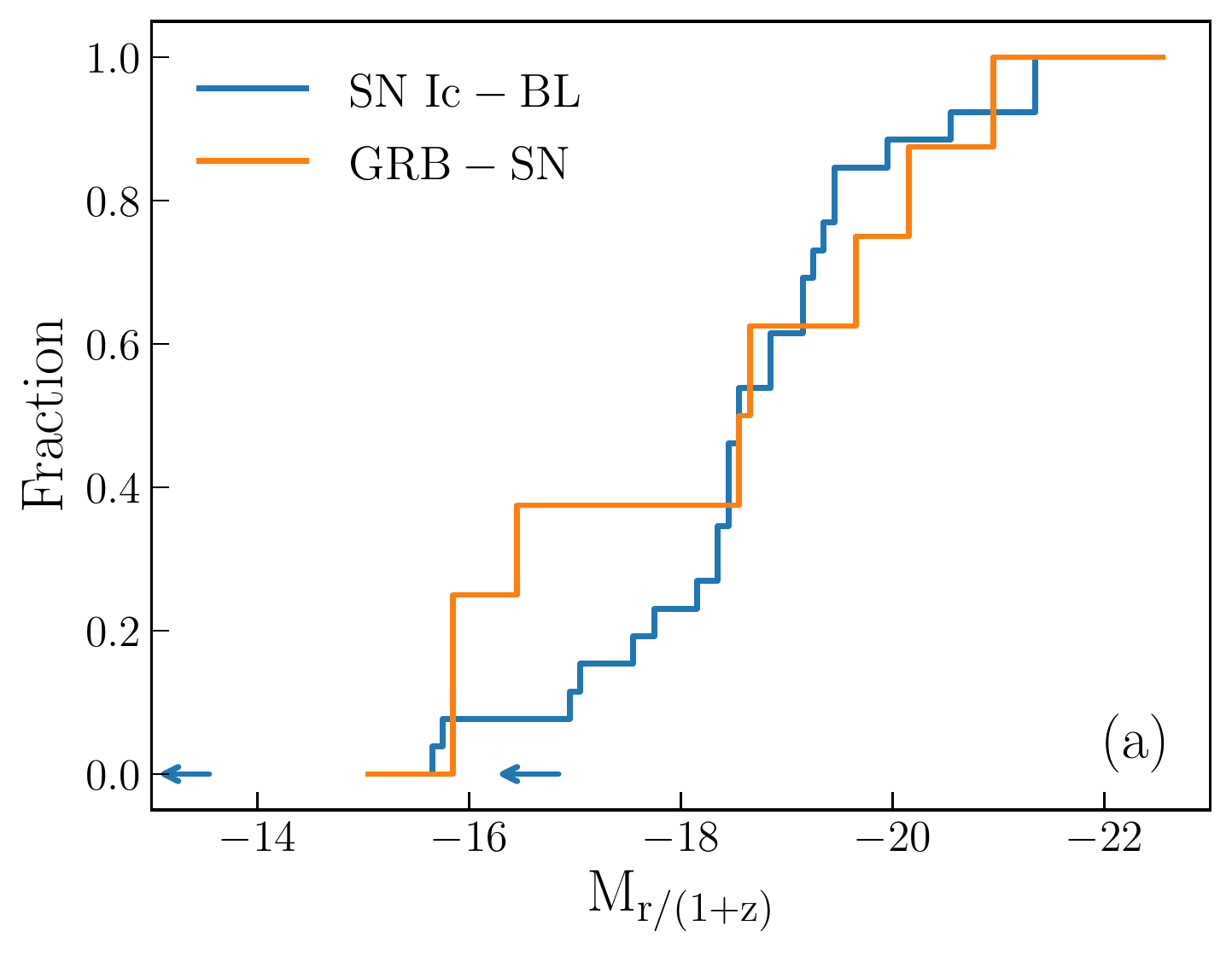}\\
\includegraphics[scale=0.55]{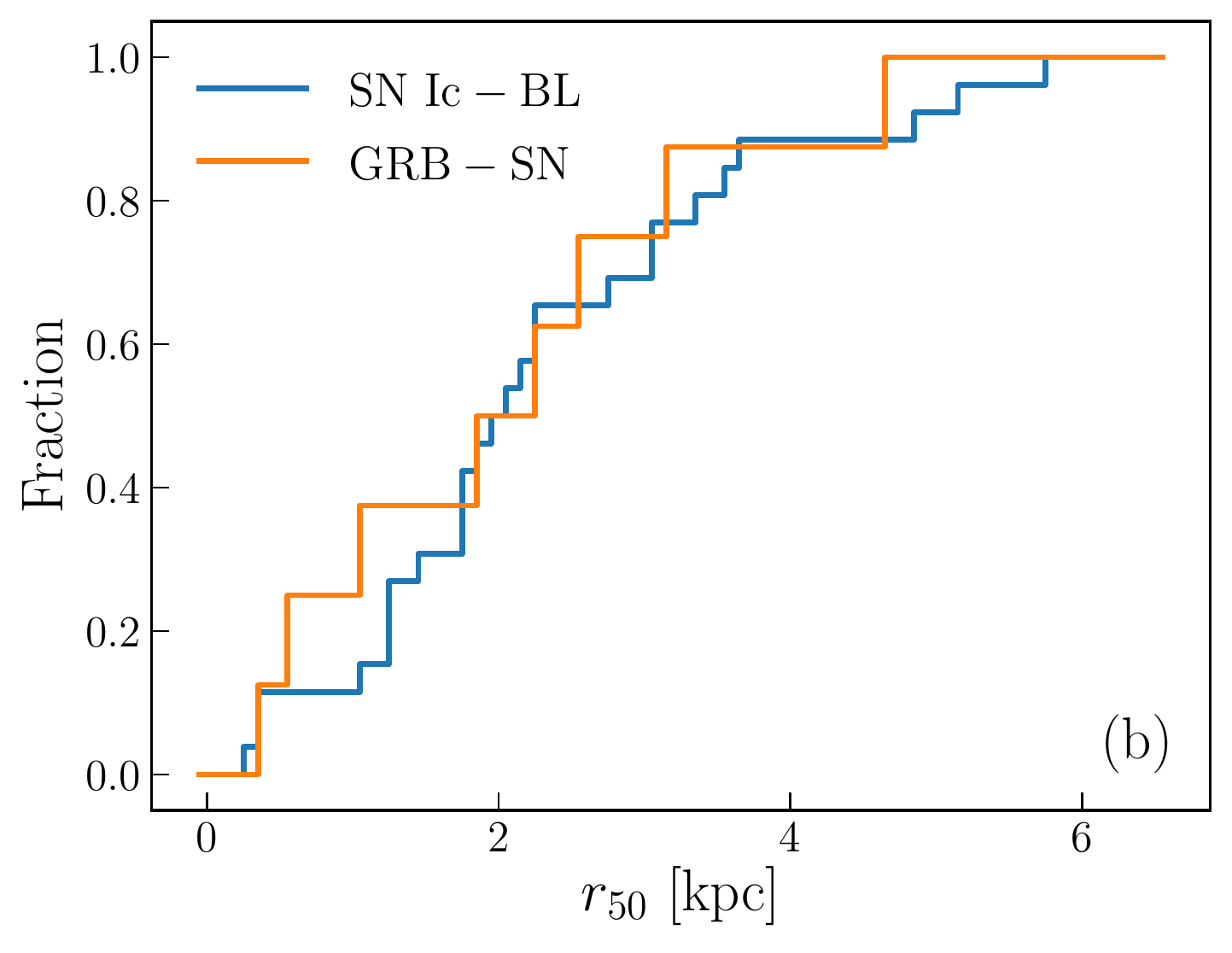}\\
\includegraphics[scale=0.55]{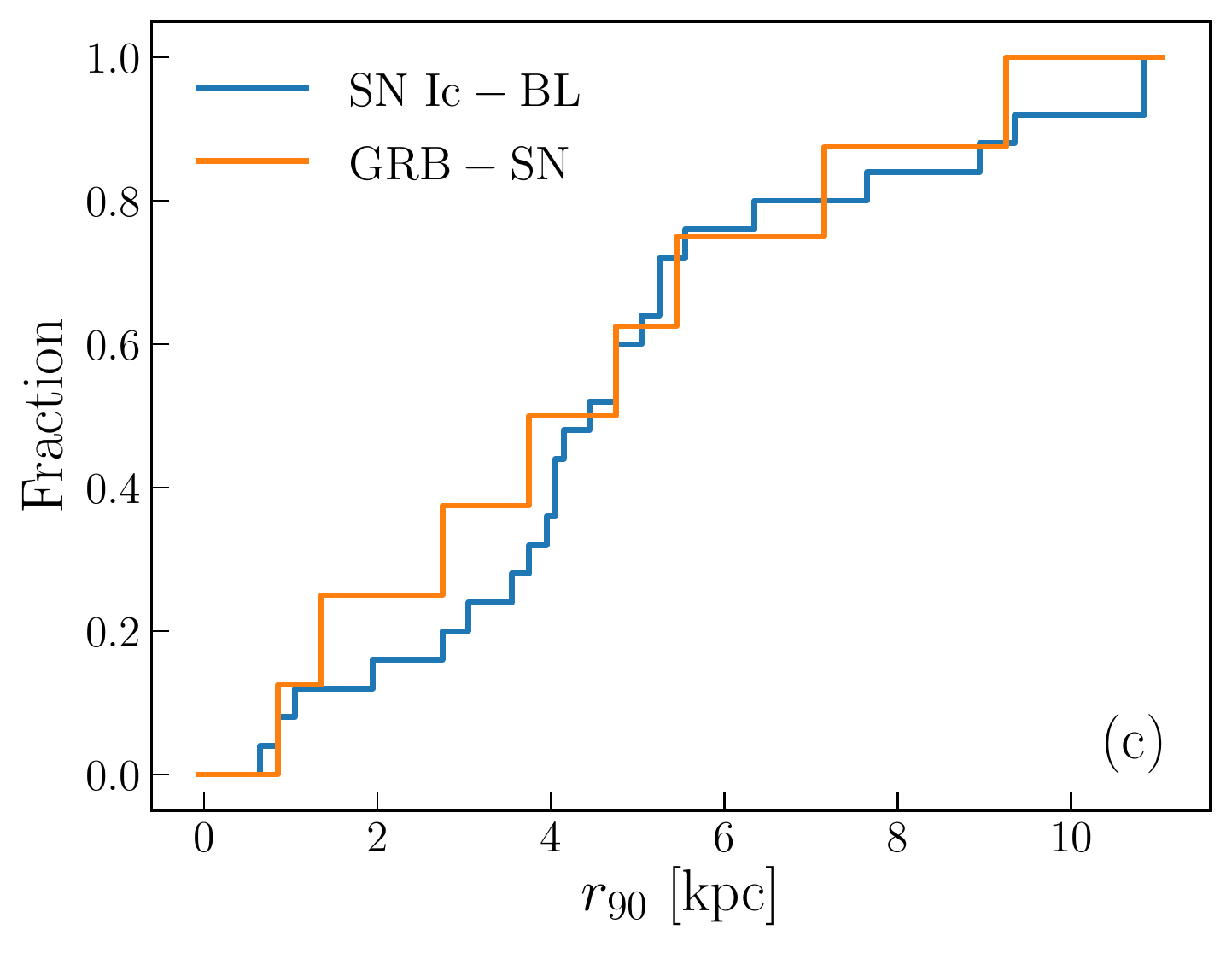}
\end{tabular}
\caption{Comparing cumulative distributions of (a) luminosity, (b) $r_{50}$ radii and (c) $r_{90}$ radii of the GRB-SN and SN Ic-BL samples. The blue arrows in figure (a) correspond to the upper limits measured for the hosts of SN2007bg and PTF11img.}
\label{fig3}
\end{figure}

\begin{table}
\small
\renewcommand{\arraystretch}{1.4}
\begin{center}
\begin{tabular}{lcccr}
\hline
\hline
                                  & \multicolumn{2}{c}{Median ($\pm1\sigma$)} & \multicolumn{2}{c}{AD test}\\
\cmidrule(lr{.75em}){2-3}\cmidrule(lr{.75em}){4-5}
Characteristic                    & SN Ic-BL & GRB-SN & $p_{\rm AD}$ & $p_{\rm AD,MC}$\\
\hline
$r_{\rm 50}$ (kpc)                & 1.99$_{-1.05}^{+1.90}$ & 2.05$_{-1.60}^{+1.94}$ & 0.87 & 0.87$_{-0.07}^{+0.11}$\\
$r_{\rm 90}$ (kpc)                & 4.58$_{-2.83}^{+5.04}$ & 4.26$_{-3.16}^{+4.02}$ & 0.87 & 0.90$_{-0.15}^{+0.13}$\\
Offset$_{\rm centre}$ (kpc)       & 3.08$_{-2.35}^{+2.98}$ & 1.54$_{-1.28}^{+3.13}$ & 0.15 & 0.18$_{-0.13}^{+0.16}$\\
Offset$_{\rm bp}$     (kpc)       & 3.15$_{-2.52}^{+3.16}$ & 0.76$_{-0.51}^{+4.26}$ & 0.10 & 0.07$_{-0.04}^{+0.09}$\\
Offset$_{\rm centre}$/r$_{\rm 50}$& 1.43$_{-0.86}^{+0.99}$ & 0.98$_{-0.74}^{+0.43}$ & 0.07 & 0.07$_{-0.04}^{+0.07}$\\
Offset$_{\rm bp}$/r$_{\rm 50}$    & 1.52$_{-0.92}^{+1.03}$ & 0.75$_{-0.52}^{+0.54}$ & 0.02 & 0.02$_{-0.01}^{+0.04}$\\
\hline
\hline
\end{tabular}
\end{center}
\caption{Statistical description of the data of the SN Ic-BL and GRB-SN samples. Reported are the median and 1$\sigma$-equivalent values of the distributions of several galaxy characteristics of our samples. The last two columns give the results of the two-sided Anderson-Darling test for comparing the distributions of the corresponding characteristics of the two samples. AD chance probabilities are given for the case when distributions are taken at face values ($p_{\rm AD}$) and when errors and systematics are taken into account ($p_{\rm AD,MC}$).}
\label{tab5}
\end{table}

The two samples have a similar absolute magnitude distribution (Fig. \ref{fig3}a), especially when considering the two non-detections. Without the non-detections the two-sided Anderson-Darling test between the two distributions gives a chance probability of $p_{\rm ch} = 0.72$ that the two distributions are drawn from the same parent distribution. In calculating absolute magnitudes we did not take into account the fact that the galaxies are at different redshifts and that they are observed with different filters. Given the redshift range and typical spectral shapes the different filter characteristics are expected to introduce uncertainties at the level of a few to $\sim 10\%$. Nevertheless, as there are many uncertainties involved in the comparison of absolute magnitudes, the results should be taken with caution and we do not include a detailed statistical summary in Table \ref{tab5}.

The galaxies in the two samples have on average similar sizes both when considering $r_{50}$ and $r_{90}$ (Fig. \ref{fig3}). This is confirmed by performing an AD test (see Table \ref{tab5}). Many studies of long GRB host galaxy samples have done similar measurements as presented in this work, mostly using HST images for the measurements. Median sizes of long GRB hosts in the literature (typically covering samples in large redshift ranges and measured on images taken at different rest-frame wavelengths) are $r_{\rm 50,med} = 1.8 \pm 0.1$ kpc \citep{Blanchard2016}, $1.7 \pm 0.2$ kpc \citep{Lyman2017} and $1.5 \pm 0.5$ kpc \citep{Bloom2002}.

The redshift distribution of the GRB-SN sample is skewed slightly towards higher redshift with respect to the regular SN sample (Fig. \ref{fig2}). According to \citet{Shibuya15} the average size evolution of star-forming galaxies can be approximated with the $\left( 1 + z\right)^{-1.2}$ relation. Using it at face value, the small difference in the average redshift of our distributions results in a negligible difference in the mean size. Any significant difference in the size between the two populations should therefore be attributed to the intrinsic difference of the populations. 

At $z \gtrsim 0.05$ the H$\alpha$ nebular emission line falls out of the filters with which the images analysed in this work have been taken. H$\alpha$ can be quite strong in star forming galaxies and therefore it could have a (small) effect on our measurements of galaxy sizes and offsets. We check how this affects our results by analysing the $z > 0.05$ part of the SN Ic-BL sample, this time analysing the {\it i}-band SDSS and PanSTARRS images. We find a very good agreement between the sizes and offsets measured from {\it r-} and {\it i-}band images. 

\subsection{Offsets}
\label{offsets}

\begin{figure}[]
\centering
\begin{tabular}{c}
\includegraphics[scale=0.58]{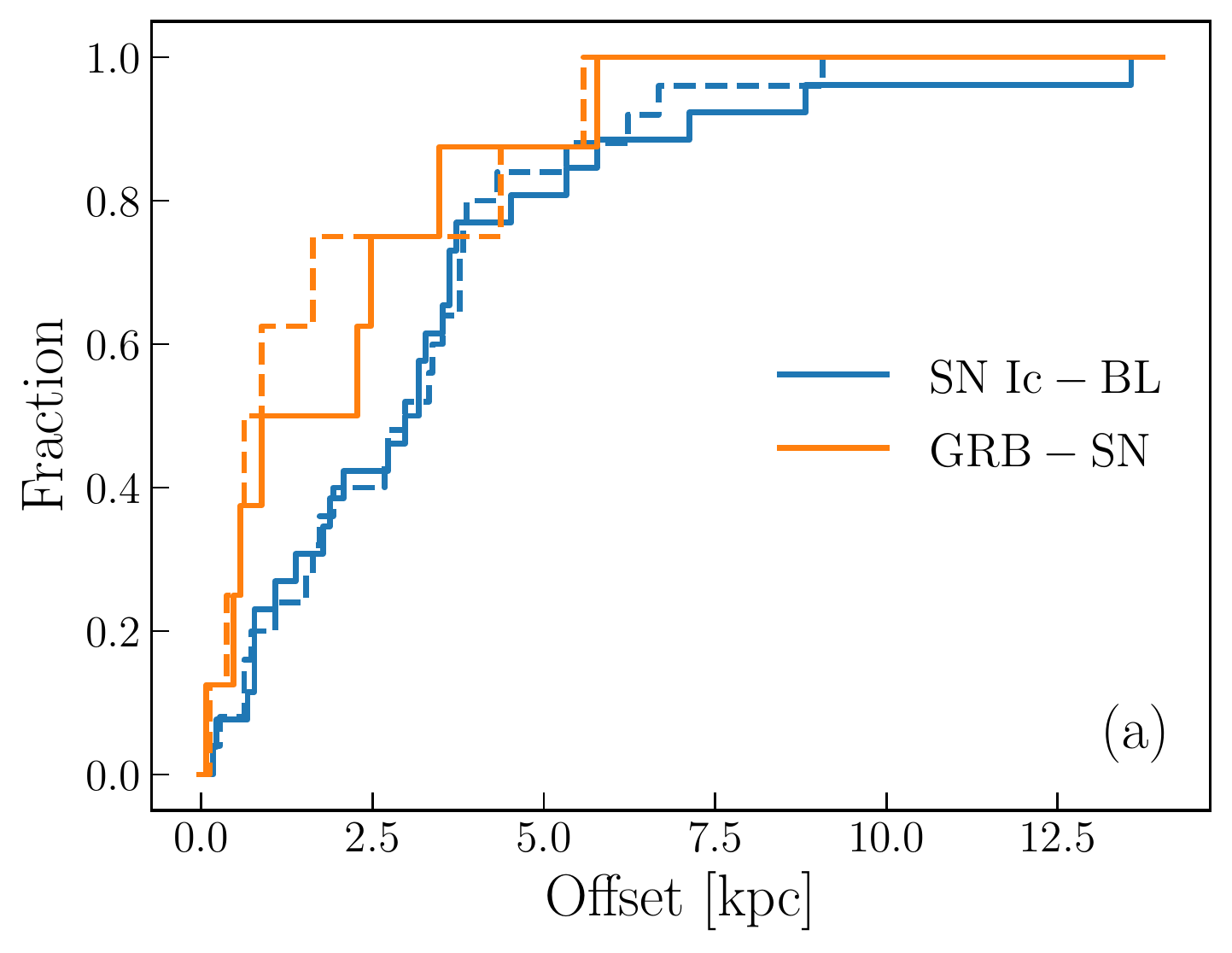}\\
\includegraphics[scale=0.58]{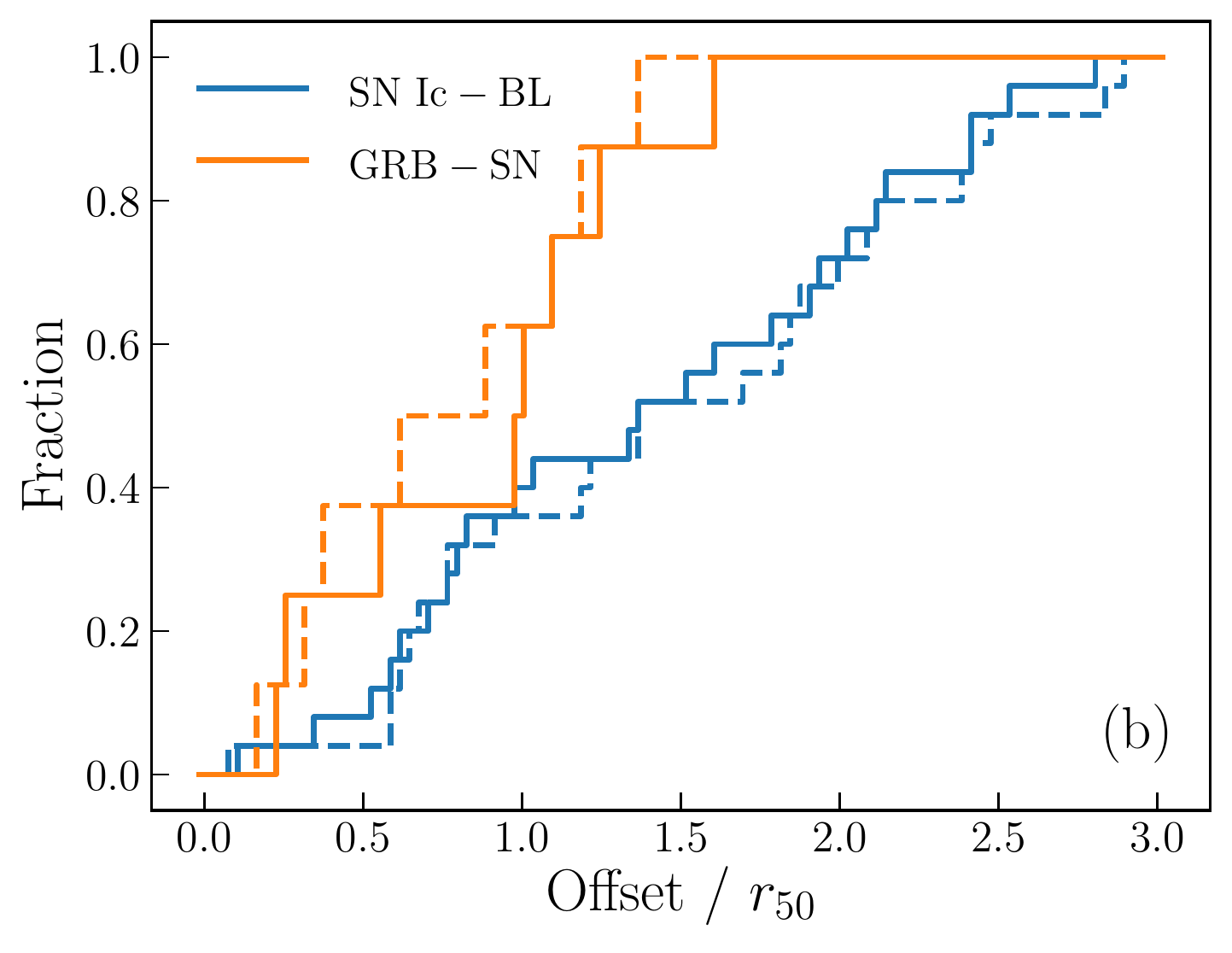}
\end{tabular}
\caption{Cumulative distributions of the projected offset (a) and the normalized offset (b) of the explosions from the host galaxy centre. Solid and dashed lines indicate offsets with respect to galaxy centres and the brightest pixel, respectively.}
\label{fig4}
\end{figure}

Next we look at the offset of the explosion from the host galaxy centre. SNe Ic-BL with GRBs are found closer to their host galaxy centre on average (Fig. \ref{fig4}a). As the galaxies span a large range in size ($r_{50} \sim 0.2 - 6$ kpc for both populations), more relevant quantity to compare is the offset normalized by the galaxy size. We use the measured $r_{50}$ for normalization. The distributions for the two samples are shown in Fig. \ref{fig4}b and they are clearly distinct, that is, GRB-SNe occur closer to their host galaxy centre than the SNe Ic-BL. This is confirmed by performing the AD test between the two distributions resulting in the chance probability of $p_{\rm ch} = 0.07$ that the two distributions are drawn from the same parent distribution. 

One can also look at the offsets from the brightest pixels. In general the position of galaxy's barycentre and its brightest pixel are similar for our samples (see Figure \ref{fig1} and Table \ref{tab2}). The difference between normalized offsets between SN positions and the brightest pixels in the galaxy is found to be even more significant ($p_{\rm ch} = 0.02$). However we note that this difference arises mostly due to the case of GRB\,980425, where the brightest pixel is not located near the galaxy's centre but instead in a well-known Wolf-Reyet region located $\approx 0.9$ kpc from the GRB \citep{Hammer2006,Kruhler17}.

The positions of SNe Ic-BL are typically reported without uncertainties. While five of the SNe discovered either by the SDSS survey (2) or the Gaia satellite (3) should be very accurate, we cannot claim that for the majority of the sample. The uncertainty in the SN position is expected to be much higher than the uncertainty in the astrometric calibration. We estimate the uncertainty of astrometric calibration of the SDSS and PanSTARRS images by comparing the positions of the relatively bright sources in the images to the catalogued values. The measured uncertainties\footnote{For typical astrometric uncertainties in the SDSS images (which agree with our estimates) see also \citet{Pier2003}.} are $<0.1\arcsec$ in all cases. Given the sizes of the galaxies in the sky, and assuming the worst case scenario in which the SN positions are uncertain by $\sim 1\arcsec$, then approximately 20$\%$ of the SN sample could be affected. In the case of GRB-SNe the positions are well constrained ($\lesssim 0.2\arcsec$) due to the availability of radio detections of bright afterglows. The measured astrometric uncertainties in the VLT/FORS2 images as well as PanSTARRS are $<0.1\arcsec$. The HST image of GRB\,030329 host was aligned with an image of the afterglow taken in the same filter. The HST image of GRB\,100316D host was aligned with an image of the afterglow taken in a similar (but not the same) filter. The estimated uncertainty for these two hosts is $<0.1\arcsec$. The astrometric calibration for the HST image of GRB\,060218 was achieved by alignment with the calibrated VLT/FORS2 image of the same field and the estimated uncertainty is $\sim 0.2\arcsec$. Combining uncertainties in astrometric calibrations with uncertainties of the afterglow position, the uncertainties are $<0.2\arcsec$ in all cases studied in this paper. For the computation of the AD test using the MC simulation we assume a conservative systematic uncertainty of $0.5\arcsec$ and $0.2\arcsec$ for SNe Ic-BL and GRB-SNe, respectively. While the computed chance significance in this case has a tale towards higher values, the results in Table \ref{tab5} still indicate a difference.

\citet{Blanchard2016} find offset$_{\rm centre,med} = 1.3 \pm 0.2$ kpc and offset$_{\rm centre}/r_{\rm 50} = 0.7 \pm 0.2$, in agreement with our results. Our results are also in agreement with \citet{Bloom2002} who find offset$_{\rm centre,med} = 1.4 \pm 0.8$ kpc and offset$_{\rm centre}/r_{\rm 50} = 0.8 \pm 0.3$, while \citet{Lyman2017} finds offset$_{\rm centre,med} = 1.0 \pm 0.2$ kpc and offset$_{\rm centre}/r_{\rm 50} = 0.6 \pm 0.1$.

In general we do not expect the galaxies to have a preferential orientation and therefore, on statistical grounds, the fact that we are comparing projected offsets should not affect our results (other than introducing additional noise to the offset distributions). However, as our samples are small, that may not necessarily be the case. We therefore estimated the inclinations of our galaxies and calculated the deprojected offsets for the galaxies of our sample (see Appendix \ref{deprojected}). The analysis shows that the trend that GRB-SNe are found closer to the centres of their host galaxies persist even when considering deprojected offsets.

The sample of GRB-SNe is small: using a bootstrap method by re-sampling the chance probabilities have a long tail towards $p_{\rm ch} > 0.1$ (while the median changes only little). In addition, the sample of SNe Ic-BL could be contaminated by actual GRB-SNe with a GRB jet being directed away from us. If one or two SNe Ic-BL with the high normalized offsets turn out to be GRB-SNe, the two distributions would become very similar (though the argument could also go in the other direction). Nevertheless, the results indicate that the two samples are drawn from different distributions, even if we cannot claim that with a high significance. With this in mind we therefore assume that the observed difference is real in the discussion that follows.

\section{Metallicities}
\label{metal}

Since metallicity is considered to be one of the main parameters that influences the lives and fates of massive stars that are considered to be the progenitors of SNe and GRBs, we searched the literature for spectroscopic analyses of the hosts included in this paper in order to measure gas-phase metallicities from nebular emission lines. \citet{Modjaz11} and \citet{Sanders12} analyzed 8 SNe Ic-BL hosts that are part of our sample. Additionally we consider 6 SNe Ic-BL hosts from \citet{Modjaz11} that were detected in a private untargeted survey, but whose classification is hard to confirm. We include these hosts in order to increase the sample - as it will be shown later this does not affect the results. For GRB-SNe, we can determine the metallicities for 6 hosts: GRB\,980425 \citep{Kruhler17}, GRB\,030329 \citep{Gorosabel05}, GRB\,031203 \citep{Guseva11}, GRB\,060218 \citep{Wiersema2007}, GRB\,100316D \citep{Izzo17}, GRB\,161219B \citep{Cano2017b}. All the host galaxies are listed in Table \ref{tab3}. For comparison we also collect spectral data for SN Ib and Ic host samples provided by \citet{Sanders12} and \citet{Modjaz11}.

Given the availability of the detected lines we compute the metallicities using two different diagnostics. The first one is the O3N2 diagnostic provided by \citealt{Pettini2004}, where the metallicity is computed from the [OIII]$\lambda5007$/H$_{\beta}$ and [NII]$\lambda6584$/H$\alpha$ ratios. In a few cases where this diagnostic could not be used we compute the metallicity using one of the other common calibrations and then transform its value to the O3N2 scale \citep{Kewley2008}. The other diagnostic is the one provided by \citet{Maiolino2008}, where the metallicity is obtained by using as many different calibrations as possible and then minimizing the probability density distribution (see also \citealt{Japelj2016,Japelj2016b} for details). This diagnostic is very similar to the calibration provided by \citet{Kewley2002}. While \citet{Modjaz11} does not provide spectral line measurements, they already compute the metallicities in both O3N2 and \citet{Kewley2002} scales, which we adopt directly in this paper. All the values and references to the source material are gathered in Table \ref{tab3}.

\begin{figure*}[]
\centering
\includegraphics[scale=0.6]{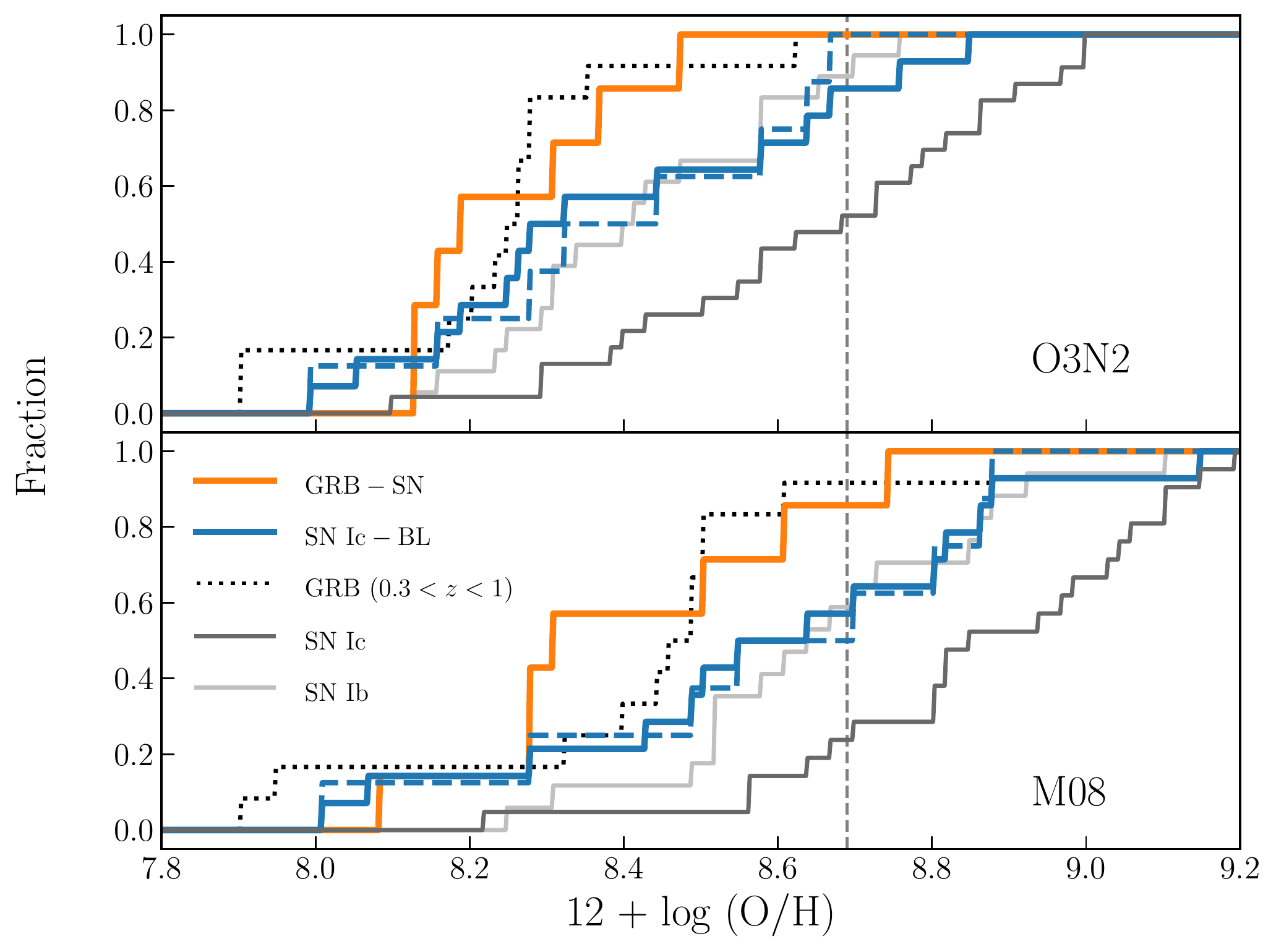}
\caption{Cumulative metallicity distributions of GRB-SN and SN Ic-BL samples: SNe Ic-BL without GRBs on average prefer environments with higher metallicities (see discussion in Section \ref{metal}). For SNe Ic-BL we plot both the distribution including all hosts with metallicities in Table \ref{tab3} (full line) as well as the distribution including only the hosts studied in this paper (dashed). The upper and lower plot show the metallicities measured using the calibrations of \citet{Pettini2004} (O3N2) and \citet{Maiolino2008} (M08), respectively. For comparison we also plot the distribution of the $z < 1$ sample of long GRB host galaxies \citep{Kruhler2015,Japelj2016} and the samples of SN Ib and Ic host galaxies from \citet{Sanders12} and \citet{Modjaz11}. The vertical dashed lines indicate solar oxygen abundance of $12 + \log ({\rm O/H}) = 8.69$ \citep{Asplund2009}.}
\label{fig5}
\end{figure*}

The metallicity distributions are plotted in Figure \ref{fig5}. For comparison we also plot the metallicity distribution of the combined $z < 1$ host galaxy complete BAT6 sample of GRBs \citep{Salvaterra2012,Japelj2016} and the sample of $z < 1$ host galaxies from \citet{Kruhler2015}. It has been shown that the metallicity distribution of GRB host galaxies does not show any evolution at least up to $z \sim 2$ \citep{Kruhler2015}, therefore the fact that the long GRB $z < 1$ sample covers different redshift regime is not problematic. Indeed, our GRB-SNe host galaxies have similar metallicities than the long GRB sample (note that the two samples do not have any galaxy in common). On the other hand, SN Ic-BL host galaxies have metallicities skewed towards higher values. It is rather striking that $\sim 30\%$ of the SNe Ic-BL are found to have super-solar metallicities. This is a much higher fraction than what is found for GRB host galaxies ($\sim 10\%$; \citealt{Kruhler2015,Japelj2016}). The difference in metallicities is observed both in the O3N2 and the \citet{Maiolino2008} diagnostics. We also note that for the SNe Ic-BL sample the resulting distribution with and without adding the additional objects from \citet{Modjaz11} is the same.

The metallicity distributions in Figure \ref{fig5} reveal that SN Ib and Ic-BL hosts have similar metallicity distributions (though the metallicities for SN Ic-BL hosts do extend to lower values). SN Ic hosts have clearly higher metallicities on average (see also \citealt{Modjaz11} and \citealt{Sanders12}). On the other hand, long GRB host galaxies seem to have lower metallicities on average than other types \citep[see also e.g.][]{Graham2013}. For the sake of completeness we provide a summary of the results of the Anderson-Darling two-sided test for metallicity distributions of different explosions' environments in Table \ref{tab4}. Based on the test, we cannot rule out that the GRB-SNe are found in the galaxies with similar metallicities as the hosts of SNe Ic-BL. If the distribution of GRB-SNe indeed follows the distribution of long GRBs, then the difference becomes slightly more significant. It is interesting to see that the lack of low-metallicity Ib hosts leads to much more significant difference to GRB-SNe and long GRB distributions, even though the distribution itself is not statistically different from the SNe Ic-BL sample. To really establish whether the hosts of these transients differ significantly or not, the number of studied hosts will have to be increased.

\subsection{Caveats}

In all but two cases the apparent galaxy size in the sky is much bigger than the slit width with which the spectrum was taken. The metallicities for these events can be considered as if being measured at the explosion sites, though the spatial resolution naturally does not match the resolutions achieved when observing very close host galaxies with integral field units like VLT/MUSE \citep[see e.g.][]{Kruhler17,Izzo17}. Metallicities for the SN2006nx and SN2007I hosts are galaxy-averaged values as the galaxies are small in the sky. Detailed studies of nearby, well resolved host galaxies of GRB-SNe with integral field spectroscopy have shown that the metallicities at the explosion sites do not differ appreciably from the host-averaged metallicities \citep[e.g.][]{Christensen2008,Levesque2011,Kruhler17,Izzo17}. The two galaxies are also physically quite small (and comparable to the sizes of GRB-SN hosts) and therefore we expect their average metallicities to be similar as the metallicity at the SN site. 

We note that the metallicities measured and quoted in this paper represent oxygen abundances. On the other hand the metallicities that are used in the computations of theoretical models are based on iron, which we cannot measure directly. In the following discussion we implicitly assume that iron scales with oxygen and the diagnostics that we use can be used as a proxy for the metal content in stars. 

\begin{table}
\small
\renewcommand{\arraystretch}{1.2}
\begin{center}
\begin{tabular}{lccr}
\hline
\hline
            & \multicolumn{2}{c}{12 + $\log \left( \frac{\rm O}{\rm H}\right)$} & \\
\cmidrule(lr{.75em}){2-3}
SN/GRB      & O3N2  & M08  & Ref \\
\hline
SN2005kr*   & 8.24 +- 0.01 & 8.63 +- 0.03 & 1\\
SN2005ks    & 8.63 +- 0.01 & 8.87 +- 0.03 & 1\\
SN2005nb    & 8.44 +- 0.03 & 8.70 +- 0.06 & 2\\
SN2006nx    & 8.28 +- 0.04 & 8.48 +- 0.06 & 2\\
SN2006qk*   & 8.75 +- 0.02 & 8.82 +- 0.05 & 1\\ 
SN2007I     & 8.58 +- 0.07 & 8.80 +- 0.07 & 2\\
SN2007ce    & 7.99 +- 0.02 & 8.00 +- 0.02 & 2\\
SN2007eb*   & 8.26 +- 0.07 & 8.43 +- 0.10 & 1\\
SN2007gx*   & 8.85 +- 0.06 & 9.14 +- 0.02 & 1\\
SN2007qw*   & 8.19 +- 0.01 & 8.50 +- 0.07 & 1\\
SN2008iu*   & 8.05 +- 0.03 & 8.06 +- 0.04 & 1\\
SN2010ah    & 8.32 +- 0.05 & 8.55 +- 0.10 & 2\\
SN2010ay    & 8.16 +- 0.02 & 8.28 +- 0.02 & 2\\
PTF10vgv    & 8.67 +- 0.06 & 8.86 +- 0.06 & 2\\
\hline
GRB\,980425  & 8.31 +- 0.01 & 8.50 +- 0.02 & 3\\
GRB\,030329  & 8.19 +- 0.10 & 8.28 +- 0.04 & 4\\
GRB\,031203A & 8.12 +- 0.06 & 8.08 +- 0.10 & 5\\
GRB\,060218  & 8.13 +- 0.03 & 8.28 +- 0.04 & 6\\
GRB\,100316D & 8.16 +- 0.01 & 8.30 +- 0.02 & 7\\
GRB\,161219B & 8.37 +- 0.08 & 8.56 +- 0.04 & 8\\
\hline
\hline
\end{tabular}
\end{center}
\caption{Metallicity measurements in the O3N2 and M08 calibrations. SNe Ic-BL tagged with * are not a part of our original sample: they are part of the untargeted survey presented by \citet{Modjaz11}, but their classification is not confirmed. References point to the source material from which we collected the measurements of emission-line fluxes. \newline References: (1) \citet{Modjaz11} (2) \citet{Sanders12} (3) \citet{Kruhler17} (4) \citet{Gorosabel05} (5) \citet{Guseva11} (6) \citet{Wiersema2007} (7) \citet{Izzo17} (8) \citet{Cano2017b}}
\label{tab3}
\end{table}

\begin{table}
\small
\renewcommand{\arraystretch}{1.2}
\begin{center}
\begin{tabular}{lcc}
\hline
\hline
            & \multicolumn{2}{c}{Chance probability $p_{\rm ch}$}\\
\cmidrule(lr{.75em}){2-3}
SN/GRB      & O3N2  & M08\\
\hline
GRB-SN - Ic-BL    & 0.43 & 0.23\\
GRB-SN - long GRB & 0.53 & 0.67\\
GRB-SN - Ic       & 0.001 & 0.0006\\
GRB-SN - Ib       & 0.07  & 0.02\\
Long GRB - Ic-BL  & 0.16  & 0.12\\
Long GRB - Ic     & 0.00001 & 0.00001\\
Long GRB - Ib     & 0.01    & 0.004\\
Ic - Ic-BL        & 0.004   & 0.004\\
Ic - Ib           & 0.004   & 0.02\\
Ib - Ic-BL        & 0.27    & 0.27\\
\hline
\hline
\end{tabular}
\end{center}
\caption{Results of the two-sided Anderson-Darling test between metallicity distributions of different SN and long GRB samples.}
\label{tab4}
\end{table}

\section{Implications for progenitors}
\label{progenitor}

Our findings potentially harbour important clues about the nature of the progenitors. In the context of the collapsar scenario \citep{Woosley1993,MacFadyen1999,MacFadyen2001} the progenitor model needs to explain how to retain enough angular momentum in the core at the moment of explosion, while losing the H- and He-rich outermost layers \citep{Woosley2006, Modjaz16}. In a single star scenario, mass loss mechanisms typically also remove angular momentum, and are sensitive to the progenitor metallicity (stellar wind mass-loss rates approximately scale as $\approx Z^{0.8}$; \citealt{Vink2001,Mokiem2007}). Conversely, binaries provide a reservoir of angular momentum stored in the orbit, which can be tapped into through tidal interactions \citep{Izzard04,Detmers08,Mink2009}, mass accretion \citep{Cantiello07,Eldridge11,Mink2013}, and/or mergers \citep{Tout11,Zapartas17a}. Moreover, binaries can increase the delay time between the star formation phase and the death of massive stars \citep{Zapartas17a}. Therefore certain binary products have on average more time to travel longer distances from their birth location, possibly resulting in larger offsets. On the other hand, if GRBs occur via a binary channel, then their preference for dense environments might actually lead to shorter merger time scales \citep[e.g.][]{Heuvel2013}.

\cite{Modjaz16} have shown that there are significant spectral differences between SNe Ic-BL with and without GRBs, which suggests two different progenitor classes, or, possibly, a continuum of progenitors resulting at the extreme in the two different phenomena. Our findings, albeit based on a limited sample of hosts, seem to corroborate this idea.

The offsets of long GRB explosions in general have first been studied by \citet{Bloom2002}, who found that GRBs indeed occur close to the brightest, star-forming regions of their hosts (see also Section 3.2). A more detailed pixel-by-pixel analysis of resolved host images have confirmed the trend and in addition revealed that GRBs are found on brighter parts of their host galaxies with respect to a general class of core-collapse SNe \citep{Fruchter2006,Kelly2008,Svensson2010,Lyman2017}. The regions of the most recent star formation are expected to be brighter than the regions where more time has passed since the star formation episode, because the massive stars have not exploded yet. This result has therefore been interpreted as evidence for massive stellar progenitors and short progenitor lifetimes of long GRBs. The difference in the distribution of offsets of the SNe Ic-BL with and without GRBs that we find (Fig. 4) implies that the progenitors of the GRB-SNe may be shorter-lived stars with respect to SNe Ic-BL. The shorter lifetime could be the natural consequence of a more massive progenitor, either in the case of single star evolution scenario or possibly resulting from stellar mergers. If the difference in the offsets indeed translates into more massive progenitors for GRB-SNe, this would be evidence against the scenario where the SNe Ic-BL without GRBs are the result of choking (damping) of the jet. In fact, we expect the choking of the jet to become progressively more efficient as the mass of the layers the jet has to plow through increases \citep[e.g.][]{Margutti14,Margutti2015,Modjaz16}. 

Above we implicitly assumed that the peak of the surface brightness traces young stellar populations. Especially at low redshifts this may not necessarily be true, e.g. in the case if the brightness is dominated by a bulge instead of a disk, in which case the high surface-brightness regions are dominated by older stars. A measurement of the concentration of light $C\equiv5\log(r_{80}/r_{20})$ \citep{Kent1985,Conselice2005} for our galaxies gives values of $C \lesssim 3$, suggesting that the light is indeed dominated by the contribution from disks and not from the bulge. Nevertheless, the fact remains that many galaxies in our sample are not really resolved and therefore we cannot measure the distances to the individual star-forming regions (as for example in the case of GRB\,980425). It is therefore more correctly to interpret the offsets as the distances to the regions with the highest stellar densities and it follows from our analysis that GRB-SNe tend to occur closer to the densest regions in their host galaxies.

Both long GRBs and GRB-SNe have a systematically lower host metallicity than SNe Ic-BL (see also \citealt{Graham2013}). The lower host metallicity of GRBs fits naturally with the expectation of reduced wind mass and angular momentum losses with lower metallicity \citep{Vink2001}. Strong stellar winds could potentially spin down the progenitor whether its high rotation is primordial or due to binary interactions. The difference in the host metallicities might be an indication of the existence of two different channels for the formation of GRB-SNe and SNe Ic-BL or evidence that a GRB jet cannot be successfully produced in the case of high metallicity.

Recently \citet{Barnes2017} performed a simulation of a jet-driven core-collapse event and demonstrated that a GRB engine can produce a long GRB and at the same time drive a SN explosion. The produced SN has characteristics typical of SN Ic-BL and its properties depend on the viewing angle. The study does not fully explore the parameter space and therefore it remains unclear whether the observed difference in the spectra of SNe Ic-BL and GRB-SNe \citep{Modjaz16} can be attributed solely to the viewing angle. The trends in the environment that we find suggest the difference is not only due to the viewing angle. It would be interesting to see how the progenitor mass and/or metallicity affect the successful break-out of the jet from the progenitor.

The metallicity distribution in Fig.\,5 shows that SNe Ic-BL have environments more similar to SNe Ib than SNe Ic. One possible explanation could be that there are different mechanisms to remove the He lines from the spectra of SNe Ic and SNe Ic-BL. However, \cite{Modjaz16} have shown that by artificially broadening the spectra of a normal SNe Ic one finds good agreement with the spectrum of a SNe Ic-BL, which is surprising if the He is removed by a different mechanism. Another possibility is that the progenitors of SNe Ic-BL are more massive than those of normal SNe Ic, and they can therefore lose their He-rich layers to winds at lower metallicity, owing to their higher luminosity that drives the wind.

Finally, while it is indisputable that metallicity is an important factor in the GRB production efficiency, it is unclear what is the effect of other properties \cite[e.g.][]{Perley2016b}. For example, \citet{Kelly14} found that GRB hosts are smaller (and more star-forming) than the hosts of SNe Ib/c and SNe II of the same stellar mass (see also \citealt{Japelj2016}). This implies that GRBs prefer dense stellar environments with a particularly intense star formation, where a higher fraction of massive stars, close binaries or interactions between stars can be produced. These results are still tentative and should be explored in more detail. We do not have sufficient photometric and spectroscopic data at this point to do a similar analysis for our samples of GRB-SN and SN Ic-BL hosts.

In the above discussion we have implicitly assumed that GRB-SNe are a subset of long GRB class. There are three known nearby long GRBs for which an associated SN has not been found even after deep follow-up observational campaigns. These are GRB\,060505 and GRB\,060614 \citep{Fynbo2006,GalYam2006,DellaValle2006,Xu2009} and GRB\,111005A \citep{Tanga17}. The long duration of GRBs 060505 and 060614 has been put into question \citep[e.g.][]{Ofek2007,Yang2015}. On the other hand, the long nature of GRB\,111005A seems to be better established. Interestingly, this GRB has been found in a metal-rich environment with low ongoing star-formation rate \citep{Tanga17}. If this class of SN-less GRBs is indeed real, then it will require a different evolutionary channel to explain it.

\section{Conclusions}
\label{conclude}

It is not yet clear whether SNe Ic-BL with and without accompanying GRB represent the same events seen at different inclination angles or if they are intrinsically different events resulting from different evolutionary channels. In order to shed light on this issue we have done a systematic analysis of the host galaxies of SNe Ic-BL with and without an associated GRB explosion. The aim of this study is to find out whether there are any systematic differences between the environments in which these two types of explosions occur. A difference in the environment would indicate that not every SN Ic-BL is accompanied by a GRB and help us to better constrain the evolutionary channels that lead to each explosion. We collected a sample of 28 untargeted SNe Ic-BL and a sample of 8 GRB-SNe at $z < 0.2$ and analysed the images of their host galaxies. We provide measurements of galaxy luminosities, galaxy sizes ($r_{50}$ and $r_{90}$) and projected offsets between the explosions and the galaxy centres. 

We find that the SN Ic-BL and GRB-SN hosts have similar luminosities and sizes. On the other hand, the projected offsets normalized by galaxy sizes of GRB-SNe seem to be skewed towards lower values with respect to the SNe Ic-BL. The difference is of low statistical significance (chance probability of $p_{\rm ch} = 0.07$ that the two distributions are drawn from the same parent distribution). Nevertheless, the result could indicate that GRB-SNe occur closer to the regions of the highest stellar densities in their host galaxies with respect to SNe Ic-BL.

We look for additional potential differences in the environments of the two classes, in particular we study the metallicity. Collecting the available spectral information of the host galaxies (i.e. values of emission-line fluxes), we estimated the gas-phase metallicities of the host galaxies of the two samples. Indeed there is a larger fraction of SN Ic-BL hosts with super-solar metallicities with respect to the host galaxies of GRB-SNe. The differences that we find suggest that the progenitors of SNe Ic-BL and GRB-SNe really differ and it is unlikely that the differences found in the spectral properties are only due to the effect of a viewing angle. It however remains to be understood whether the two types of explosions are produced through a similar or different evolutionary channel.

The progenitor star models for these explosions are still very uncertain and poorly constrained. Every observational constraint is helpful in testing, tightening and improving them. The results presented in this paper show the potential of environment studies to provide such constraints. The sample of nearby, spectroscopically confirmed GRB-SNe is small and can hardly be increased. On the other hand, the sample of SNe Ic-BL is quite large but only $\sim 35\%$ of their host galaxies have been observed spectroscopically. By increasing the number of available spectra and therefore the number of hosts with measured metallicities (and other properties like star formation rates, ages of stellar populations, etc.) the properties of these two classes of explosions will be much better constrained.

\begin{acknowledgements}

We are grateful to Enrico Ramirez-Ruiz for fruitful discussion. JJ and LK acknowledge support from NOVA and NWO-FAPESP grant for advanced instrumentation in astronomy. SDV acknowledges the support of the French National Research Agency (ANR) under contract ANR-16-CE31-0003 BEaPro (P.I.: SDV). SdM has received funding under the European Unions Horizon 2020 research and innovation programme from the European Research Council (ERC) (Grant agreement No. 715063).\\ 

Funding for SDSS-III has been provided by the Alfred P. Sloan Foundation, the Participating Institutions, the National Science Foundation, and the U.S. Department of Energy Office of Science. The SDSS-III web site is $http://www.sdss3.org/$.

SDSS-III is managed by the Astrophysical Research Consortium for the Participating Institutions of the SDSS-III Collaboration including the University of Arizona, the Brazilian Participation Group, Brookhaven National Laboratory, Carnegie Mellon University, University of Florida, the French Participation Group, the German Participation Group, Harvard University, the Instituto de Astrofisica de Canarias, the Michigan State/Notre Dame/JINA Participation Group, Johns Hopkins University, Lawrence Berkeley National Laboratory, Max Planck Institute for Astrophysics, Max Planck Institute for Extraterrestrial Physics, New Mexico State University, New York University, Ohio State University, Pennsylvania State University, University of Portsmouth, Princeton University, the Spanish Participation Group, University of Tokyo, University of Utah, Vanderbilt University, University of Virginia, University of Washington, and Yale University. \\

The Pan-STARRS1 Surveys (PS1) and the PS1 public science archive have been made possible through contributions by the Institute for Astronomy, the University of Hawaii, the Pan-STARRS Project Office, the Max-Planck Society and its participating institutes, the Max Planck Institute for Astronomy, Heidelberg and the Max Planck Institute for Extraterrestrial Physics, Garching, The Johns Hopkins University, Durham University, the University of Edinburgh, the Queen's University Belfast, the Harvard-Smithsonian Center for Astrophysics, the Las Cumbres Observatory Global Telescope Network Incorporated, the National Central University of Taiwan, the Space Telescope Science Institute, the National Aeronautics and Space Administration under Grant No. NNX08AR22G issued through the Planetary Science Division of the NASA Science Mission Directorate, the National Science Foundation Grant No. AST-1238877, the University of Maryland, Eotvos Lorand University (ELTE), the Los Alamos National Laboratory, and the Gordon and Betty Moore Foundation.\\

We acknowledge ESA Gaia, DPAC and the Photometric Science Alerts Team (http://gsaweb.ast.cam.ac.uk/alerts).\\

This research made use of Astropy, a community-developed core Python package for Astronomy (Astropy Collaboration, 2013).

\end{acknowledgements}

\bibliographystyle{aa}
\bibliography{ms_snbl_bib}
\appendix
\renewcommand{\thesection}{A.\arabic{section}}
\section{Deprojected offsets}
\label{deprojected}

We calculate deprojected offsets of explosion sites from the galaxy centres by accounting for the position angle (PA) and the inclination angle under which we view the galaxies. We estimate inclinations by assuming that galaxies can be represented as oblate spheroids \citep{Holmberg1958}:
\begin{equation}
\cos i = \frac{\left(b/a\right)^2 - q^2}{1 - q^2},
\end{equation}
where {\it i} is an inclination angle ($i = 0^{\degree}$ and 90$^{\degree}$ correspond to a galaxy looked face-on and edge-on, respectively), $b/a$ is the measured axial ratio and $q$ the intrinsic axial ratio. The value of PA and $b/a$ are obtained from our Galfit modelling - in case the galaxy is modelled with two components with significantly different values, then PA and $b/a$ of the more extended component are assumed for the measurement of the deprojected offset. We assume a typical intrinsic axial ratio of $q = 0.2$ \citep[see e.g.][]{Unterborn2008}. If the measured axial ratio $b/a < q$, then we discard the galaxy from further analysis. This condition depends on our choice for the value of $q$. Note that the uncertainty in the measurement of inclination using this method is especially high at high inclinations (i.e. low $b/a$): it is indeed better to discard these systems in further analysis as the introduced uncertainty can be very high. If a galaxy is of irregular morphology (or the $b/a$ and PA values are poorly constrained in the fit) we assume that the galaxy is spherical and we do not apply the deprojection.

Once the inclination and position angles are known, we apply the deprojection. We firstly rotate the coordinate system for the PA, so that the long axis of the galaxy is parallel to the $y$-axis, then we apply the deprojection ($y/\cos i$) and calculate the deprojected offset. The values are given in Table \ref{tab6}. 

\begin{table}
\tiny
\renewcommand{\arraystretch}{1.2}
\begin{center}
\begin{tabular}{lcccr}
\hline
\hline
SN/GRB         & b/a  & PA & i & offset$_{depr}$\\
               &      & (deg) & (deg) & (kpc) \\
\hline
SN2005ks       & $0.31 \pm 0.05$ & $-10 \pm 4$ & 76 & 2.12\\
SN2005nb       & $0.41 \pm 0.02$ & $38 \pm 3$  & 69 & 6.14\\
SDSS-IISN14475 & $0.11 \pm 0.18$ & $-5 \pm 10$ & -  & *   \\
SN2006nx       & $0.74 \pm 0.04$ & $75 \pm 10$ & 43 & 5.69\\
SN2007bg       & -               & -           & -  & -   \\
SN2007ce$^{a}$ & $<1$            & $58 \pm 13$ & 0  & 0.65$^{\dagger}$\\
               & $0.72 \pm 0.04$ & $10 \pm 3$  & 45 & 4.00\\
SN2007I        & $0.61 \pm 0.01$ & $-36 \pm 2$ & 54 & 1.68\\
PTF10aavz      & $0.56 \pm 0.05$ & $76 \pm 2$  & 56 & 5.57\\
SN2010ah       & $0.14 \pm 0.01$ & $7 \pm 1$  & -  & *   \\
SN2010ay       & $0.61 \pm 0.06$ & $60 \pm 5$  & 54 & 0.24\\
PTF10qts       & $0.79 \pm 0.15$ & $49 \pm 30$ & 40 & 2.07\\
PTF10vgv       & $0.45 \pm 0.02$ & $45 \pm 2$  & 66 & 2.02\\
PTF10xem       & $0.48 \pm 0.03$ & $-40\pm 5$  & 64 & 6.51\\
PTF11cmh       & $0.73 \pm 0.05$ & $-52 \pm 10$& 44 & 5.82\\
PTF11lbm       & $0.20 \pm 0.02$ & $-71 \pm 1$ & - & *\\
PTF12as        & $0.79 \pm 0.03$ & $-78 \pm 30$& 39 & 3.65\\
PTF13alq       & $< 0.15$        & $45 \pm 25$ & 0  & 0.90$^{\dagger}$\\
PTF13ebw       & $0.45 \pm 0.02$ & $-27 \pm 2$ & 66 & 10.92\\
PTF13u         & $0.83 \pm 0.01$ & $37 \pm 7$  & 35 & 13.86\\
LSQ14bef       & $0.24 \pm 0.01$ & $-78 \pm 2$ & 82 & 8.08\\
PTF14dby       & $< 1$           & $50 \pm 140$& 0  & 0.75$^{\dagger}$\\
PTF14gaq       & $0.62 \pm 0.05$ & $-85 \pm 7$ & 53 & 5.75\\
iPTF15dld$^{a}$& $< 1$           & $-85 \pm 151$& 0 & 0.20$^{\dagger}$\\
               & $0.17 \pm 0.02$ & $-42 \pm 2$ & -  & *\\
SN2016coi      & $0.76 \pm 0.02$ & $-42 \pm 4$ & 42 & 2.95\\
SN2016dst      & -               & -           & 0  & 1.08$^{\dagger}$\\
SN2017dcc      & $0.58 \pm 0.01$ & $-78 \pm 3$ & 56 & 4.06\\
SN2017fgk      & $0.30 \pm 0.01$ & $61 \pm 1$  & 77 & 0.82\\
\hline
GRB\,980425    & $0.84 \pm 0.02$ & $-14 \pm 2$ & 34 & 2.33\\
GRB\,030329    & $0.77 \pm 0.05$ & $-82 \pm 3$ & 39 & 0.93\\
GRB\,031203A   & $0.83 \pm 0.03$ & $87 \pm 5$  & 35 & 0.62\\
GRB\,060218    & $< 1$           & $46 \pm 30$ & 0  & 0.08$^{\dagger}$\\
GRB\,100316D   & -               & -           & 0  & 2.48$^{\dagger}$\\
GRB\,130702A   & $0.54 \pm 0.02$ & $20 \pm 3$  & 59 & 0.92\\
GRB\,161219B   & $0.11 \pm 0.05$ & $26 \pm 3$  & -  & * \\
GRB\,171205A   & $0.81 \pm 0.01$ & $-62 \pm 5$ & 43 & 7.82\\
\hline
\hline
\end{tabular}
\end{center}
\caption{Geometric properties of the host galaxies: the axial ratio $b/a$, the position angle $PA$ (measured with respect to the North in the counter clockwise direction) and the inclination $i$. The calculated deprojected offset (with respect to the centre of the galaxy) is given in the final column. We do not provide errors for $i$ and offset$_{depr}$ due to a large uncertainties introduced in the measurement of the former.\newline$*$ Galaxy is not included in the deprojection study because $b/a < 0.2$. \newline$\dagger$ Galaxy assumed to be a spheroid observed at $i = 0$, i.e. no deprojection is applied to the offset. \newline(a)The first line corresponds to the measurement assuming the host galaxy is only the bright (blue) region near the SN explosion. The second line is obtained in the case the host galaxy is the whole complex.}
\label{tab6}
\end{table}

\begin{figure*}[t]
\centering
\begin{tabular}{cc}
\includegraphics[scale=0.58]{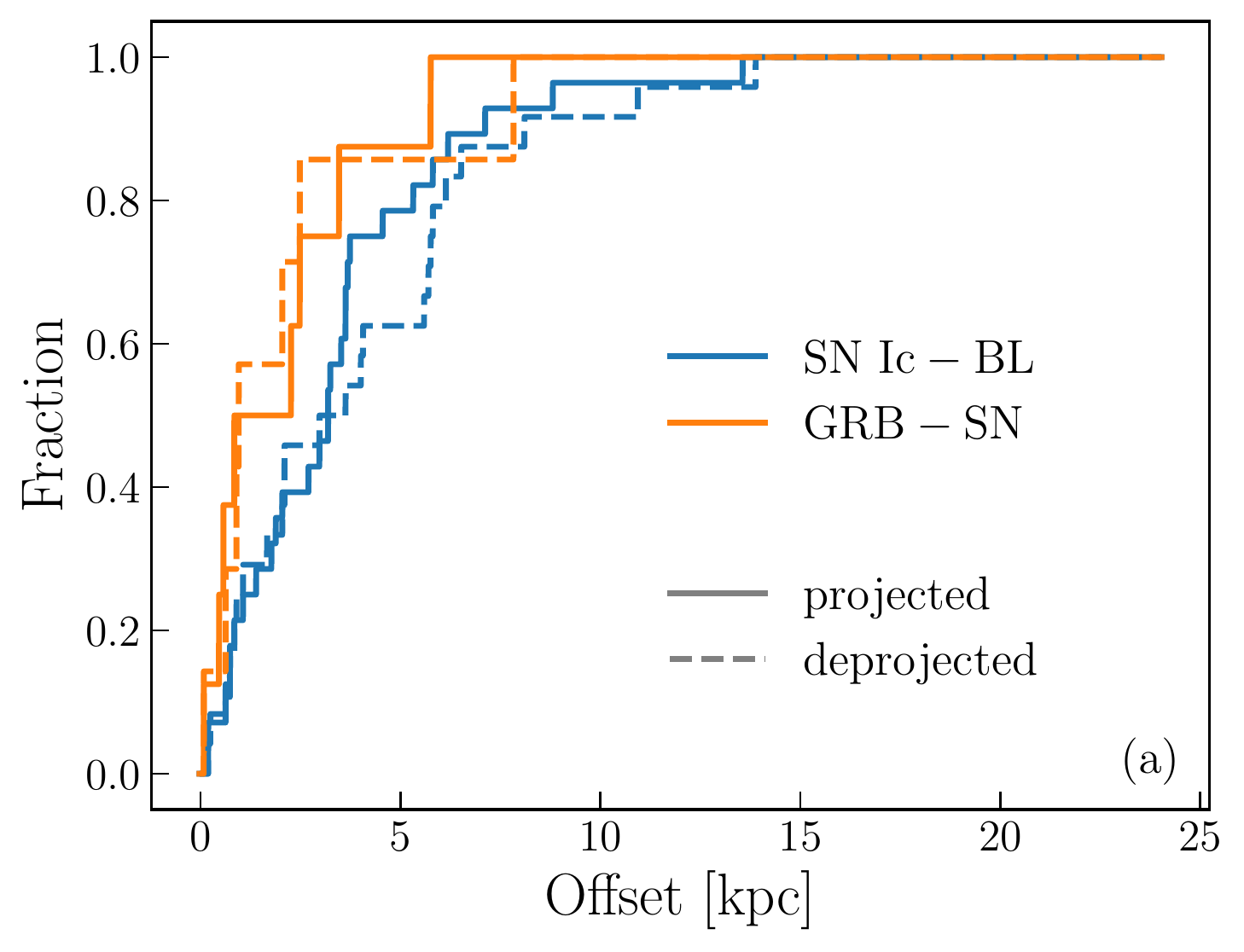}&
\includegraphics[scale=0.58]{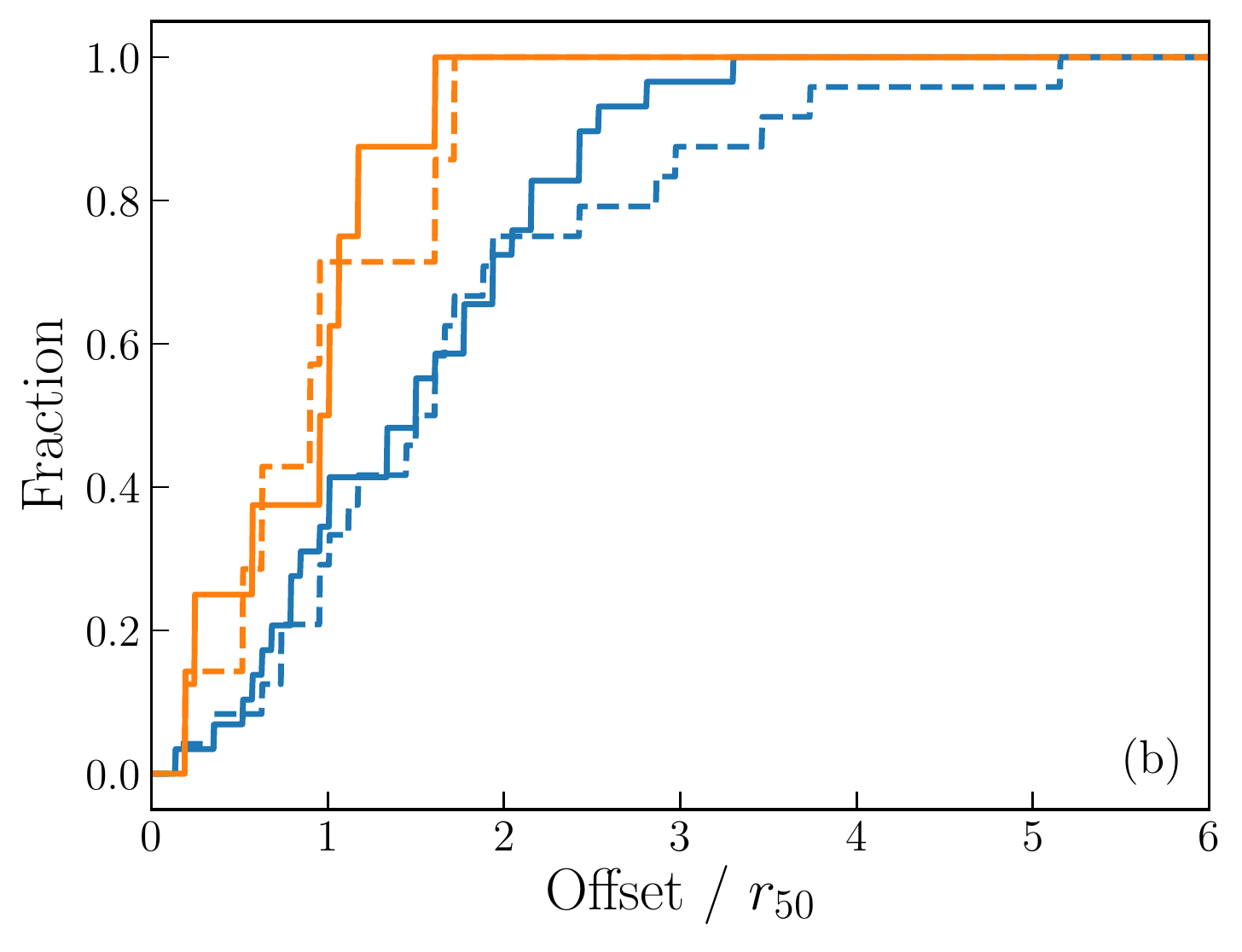}\\
\end{tabular}
\caption{Comparing cumulative distributions of (a) offsets and (b) normalized offsets of the SN Ic-BL and GRB-SN samples. Dashed lines show distributions of deprojected offsets, where the geometric properties of the galaxies (see Table \ref{tab6}) have been taken into account.}
\label{figA1}
\end{figure*}

In Figure \ref{figA1} we compare the projected and deprojected offsets of our two samples. The trend that GRB-SNe are found closer to their host galaxy centres persists when deprojected offsets are considered. The AD test gives the chance probability $p_{\rm AD} = 0.25$ and 0.1 that the distributions of the deprojected offsets and the normalized deprojected offsets are the same, respectively. Given the large uncertainties in the measurements of inclinations we do not perform a more detailed MC simulation to estimate the chance probabilities. We note that the assumption we make for the value of $q$ may not give very accurate values of inclination for individual galaxies but should work well for large samples. Given that our samples (especially the GRB-SN sample) are small and that we are analysing a selected sample of galaxies this might introduce a systematic uncertainty into our measurements. 

\end{document}